\documentclass[12pt,draftclsnofoot,onecolumn,journal]{IEEEtran}

\pdfoutput=1 

\IEEEoverridecommandlockouts 

\usepackage[cmex10]{amsmath}
\usepackage{graphicx}
\usepackage{cite}
\usepackage{array}
\usepackage{url}
\usepackage{cases}
\usepackage{amsfonts,amssymb}
\usepackage{booktabs}

\ifCLASSOPTIONcompsoc
\usepackage[caption=false,font=normalsize,labelfont=sf,textfont=sf]{subfig}
\else
\usepackage[caption=false,font=footnotesize]{subfig}
\fi

\graphicspath{{./figures/}}

\DeclareMathAlphabet{\CMmathcal}{OMS}{cmsy}{m}{n}
\renewcommand{\mathcal}[1]{\CMmathcal{#1}}
\newcommand{\figref}[1]{Fig.~\ref{#1}}
\newcommand{\secref}[1]{Sec.~\ref{#1}}

\usepackage{cite}
\usepackage{setspace}
\newcommand{\comment}[1]{{{}}}

\newcommand{\eqtop}[1]{\overset{(#1)}{=}}

\newcommand{\vect}[1]{\mathbf{#1}}

\newtheorem{proposition}{Proposition}
\newtheorem{corollary}{Corollary}
\newtheorem{remark}{Remark}

\newcommand*{\QEDA}{\hfill\IEEEQED}%

\begin{document}

\title{Energy Efficiency and Sum Rate Tradeoffs for Massive MIMO Systems with Underlaid Device-to-Device Communications}

\author{Serveh~Shalmashi,~\IEEEmembership{Student~Member,~IEEE}, Emil~Bj\"ornson,~\IEEEmembership{Member,~IEEE}, Marios~Kountouris,~\IEEEmembership{Senior Member,~IEEE}, Ki~Won~Sung,~\IEEEmembership{Member,~IEEE}, and M\'erouane~Debbah,~\IEEEmembership{Fellow,~IEEE}%
    \thanks{Serveh~Shalmashi and Ki~Won~Sung are with the Dept.\ of Communication Systems, KTH Royal Institute of Technology, Stockholm, Sweden (emails: \{serveh,sungkw\}@kth.se).}%
      \thanks{Emil~Bj\"ornson is with the Dept.\ of Electrical Engineering (ISY), Link\"oping University, Link\"oping, Sweden,  (email: emil.bjornson@liu.se).}%
      \thanks{Marios~Kountouris and M\'erouane~Debbah are with the Mathematical and Algorithmic Sciences Lab, France Research Center, Huawei Technologies Co. Ltd.  (emails: \{marios.kountouris,merouane.debbah\}@huawei.com).}%
 \thanks{Part of the material in this paper will be presented at IEEE International Conference on Communications (ICC) Workshop on Device-to-Device Communication for Cellular and Wireless Networks, London, UK, June 2015 \cite{Shalmashi-2015-ICC}.}
 }

\maketitle

\begin{abstract}
In this paper, we investigate the coexistence of two technologies that have been put forward for the fifth generation (5G) of cellular networks, namely, network-assisted device-to-device (D2D) communications and massive MIMO (multiple-input multiple-output). Potential benefits of both technologies are known individually, but the tradeoffs resulting from their coexistence have not been adequately addressed. To this end, we assume that D2D users reuse the downlink resources of cellular networks in an underlay fashion. In addition, multiple antennas at the BS are used in order to obtain precoding gains and simultaneously support multiple cellular users using multiuser or massive MIMO technique. Two metrics are considered, namely the average sum rate (ASR) and energy efficiency (EE). We derive tractable and directly computable expressions and study the tradeoffs between the ASR and EE as functions of the number of BS antennas, the number of cellular users and the density of D2D users within a given coverage area. Our results show that both the ASR and EE behave differently in scenarios with low and high density of D2D users, and that coexistence of underlay D2D communications and massive MIMO is mainly beneficial in low densities of D2D users.
\end{abstract}


\section{Introduction}

\bstctlcite{IEEEexample:BSTcontrol} 

The research on future mobile broadband networks, referred to as the fifth generation (5G), has started in the past few years. In particular, stringent key performance indicators (KPIs) and tight requirements have been introduced in order to handle higher mobile data volumes, reduce latency, increase the number of connected devices and at the same time increase the energy efficiency (EE) \cite{Osseiran-2014-COMM, Bjornson-2014-b-SPM}. The current network and infrastructure cannot cope with 5G requirements---fundamental changes are needed to handle future non-homogeneous deployments as well as new trends in user behavior such as high quality video streaming and future applications like augmented reality. 5G technology is supposed to evolve existing networks and at the same time integrate new dedicated solutions to meet the KPIs \cite{Bjornson-2014-b-SPM}. The new key concepts for 5G include massive MIMO (multiple-input multiple-output), ultra dense networks (UDN), device-to-device (D2D) communications, and huge number of connected devices, known as machine-type communications (MTC). The potential gains and properties of these different solutions have been studied individually, but the practical gains when they coexist and share network resources are not very clear so far. In this paper, we study the coexistence of two of these main concepts, namely massive MIMO and D2D communication.

Massive MIMO is a type of multiuser MIMO (MU-MIMO) technology where the base station (BS) uses an array with hundreds of active antennas to serve tens of users on the same time/frequency resources by coherent transmission processing \cite{Marzetta2010a,Rusek-2013-SPM}. Massive MIMO techniques are particularly known to be very spectral efficient, in the sense of delivering high sum rates for a given amount of spectrum \cite{Bjornson2016a}. This comes at the price of deploying more transceiver hardware, but the solution is still likely to improve the energy efficiency of networks \cite{Ngo2013a,Bjornson-2014-arxiv}.
On the other hand, in a D2D communication, user devices can communicate directly with each other and the user plane data is not sent through the BS \cite{Doppler-2009-COMM}. D2D communications are considered for close proximity applications which have the potential to achieve high data rates with little amount of transmission energy, if interference is well-managed. In addition, D2D communications can be used to decrease the load of the core network. D2D users either have their own dedicated time/frequency resources (overlay approach) which in turn leads to elimination of the cross-tier interference between the two types of users (i.e., cellular and D2D users), or they transmit simultaneously with cellular users in the same resource (underlay approach).

We consider two network performance metrics in this work: The average sum rate (ASR) in $\mathrm{bit/s}$ and the EE which is defined as the number of bits transmitted per Joule of energy consumed by the transmitted signals and the transceiver hardware. It is well-known that these metrics depend on the network infrastructure, radio interface, and underlying system assumptions \cite{Tombaz-2011-WCM, Bjornson-2014-arxiv, Auer-2013-WCM}.
The motivation behind our work is to study how the additional degrees of freedom resulting from high number of antennas in the BS can affect the ASR and EE of a multi-tier network where a D2D tier is bypassing the BS, and how a system with massive MIMO is affected by adding a D2D tier. We focus on the downlink since majority of the payload data and network energy consumption are coupled to the downlink \cite{Tombaz-2011-WCM}. We assume that each D2D pair is transmitting simultaneously with the BS in an underlay fashion. In addition, we assume that the communication mode of each user (i.e., D2D or cellular mode) has already been decided by higher layers.

\subsection{Related Work}

The relation between the number of BS antennas, ASR and EE in cellular networks has been studied in \cite{Ngo2013a, Yang-2013-OnlineGreenCom, Bjornson-2014-arxiv, Bjornson-2013-ICT} among others. The tradeoff between ASR and EE was described in \cite{Ngo2013a} for massive MIMO systems with negligible circuit power consumption. This work was continued in \cite{Yang-2013-OnlineGreenCom} where radiated power and circuit power were considered. In \cite{Bjornson-2014-arxiv}, joint downlink and uplink design of a cellular network was studied in order to maximize EE for a given coverage area. The maximal EE was achieved by having a hundred BS antennas and serving tens of users in parallel, which matches well with the massive MIMO concept. Furthermore, the study \cite{Bjornson-2013-ICT} considered a downlink scenario in which a cellular network has been overlaid by small cells. It was shown that by increasing the number of BS antennas, the array gain allows for decreasing the radiated signal energy while maintaining the same ASR. However, the energy consumed by the transceiver chains increases. Maximizing the EE is thus a complicated problem where several counteracting factors need to be balanced. This stands in contrast to maximization of the ASR, which is relatively straightforward since the sum capacity is the fundamental upper bound.

There are only a few works in the D2D communication literature where the base stations have multiple antennas \cite{Min-2011-TWC-b, Yu-2012-GLOBECOM, Fodor-2011-GLOBECOM,Shalmashi-2014a-WCNC, Xingqin-2014-arxiv}.
In \cite{Min-2011-TWC-b}, uplink MU-MIMO with one D2D pair was considered. Cellular user equipments (CUEs) were scheduled if they are not in the interference-limited zone of the D2D user. The study \cite{Yu-2012-GLOBECOM} compared different multi-antenna transmission schemes. In \cite{Fodor-2011-GLOBECOM}, two power control schemes were proposed for a multi-cell MIMO network. Two works that are more related to our work are \cite{Shalmashi-2014a-WCNC} and \cite{Xingqin-2014-arxiv}. The former investigates the mode selection problem in the uplink of a network with potentially many antennas at the BS. The impact of the number of antennas on the quality-of-service and transmit power was studied when users need to decide their mode of operation (i.e., D2D or cellular). The latter study, \cite{Xingqin-2014-arxiv}, only employs extra antennas in the network to protect the CUEs from interference of D2D users in the uplink.

The ASR in D2D communications is mostly studied in the context of interference and radio resource management \cite{Shalmashi-2013-PIMRC, Zulhasnine-2010-WiMob}. There are a few works that consider EE in D2D communications, but only for single antenna BSs, e.g.,  \cite{Yaacoub-2012-GCW, Mumtaz-2014-ICC}, and \cite{Wang-2013-ICC}, where the first one proposed a coalition formation method, the second one  designed a resource allocation scheme, and the third one aimed at prolonging the battery life of user devices.

The spatial degrees of freedom offered by having multiple antennas at BSs are very useful in the design of future mobile networks, because the spatial precoding enables dense multiplexing of users while keeping the inter-user interference under control. In particular, the performance for cell edge users, which have almost equal signal-to-noise ratios (SNRs) to several BSs, can be greatly improved since only the desired signals are amplified by  the transmit precoding \cite{Baldemair-2013-VTM,Bjornson2013d,Gesbert-2007-SPM}. In order to model the random number of users and random user positions, we use mathematical tools from stochastic geometry \cite{Haenggi-Stochastic} which are  powerful in analytically quantifying certain metrics in closed-form.

\subsection{Contributions}
Our main contributions in this paper can be summarized as follows:
\begin{itemize}
\item
\emph{A tractable model for underlaid D2D communication in massive MIMO systems}: We model a two-tier network with two different user types. The first tier users, i.e., CUEs, are served in the downlink by a BS using massive multiuser MIMO precoding to cancel interference. The second tier users, i.e., D2D users, exploit their close proximity and transmit simultaneously with the downlink cellular transmissions bypassing the BS. The number of D2D transmitters and their locations are modeled according to a homogeneous Poisson point process (PPP) while a fixed number of CUEs are randomly distributed in the network.
\item
\emph{Tractable and directly computable expressions}:
We derive tightly approximated expressions for the coverage probability of D2D users and CUEs. These expressions are directly used to compute our main performance metrics, namely, the ASR and EE. We verify the tightness of these approximations by Monte-Carlo simulations. Furthermore, we provide analytical insights on the behavior of these metrics for both CUEs and D2D users.

To the best of our knowledge, the energy efficiency analysis for underlay D2D communications in a network with large number of BS antennas has not been carried out before.
\item
\emph{Performance analysis}:
Based on extensive simulations, we characterize the typical relation between the ASR and EE metrics in terms of the number of BS antennas, the number of CUEs, and the D2D user density for a given coverage area and study the incurred tradeoffs in two different scenarios.
\end{itemize}

\section{System Model}
\label{sec:Sys_Mod}

\begin{figure}[t]
\centering
\includegraphics[width=0.6\columnwidth]{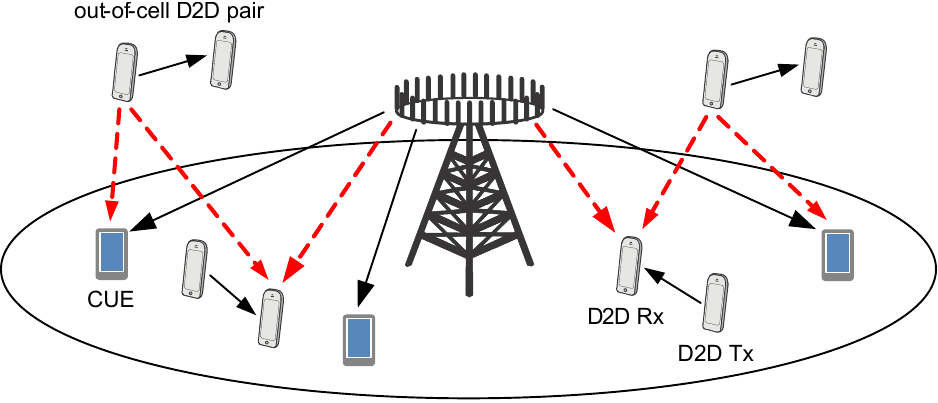}
\caption{System model where a multi-antenna BS communicates in the downlink with multiple CUEs, while multiple user pairs communicate in D2D mode. The CUEs are distributed uniformly in the coverage area and the D2D users are distributed according to a PPP. The D2D users that are outside the coverage area are only considered as interferers.}
\label{figure:sysMod_massiveMIMO_D2D}
\end{figure}

We consider a single-cell scenario where the BS is located in the center of the cell and its coverage area is a disc of radius $R$. The BS serves $U_c$ single-antenna CUEs which are uniformly distributed in the coverage area. These are simultaneously served in the downlink using an array of $T_c$ antennas located at the BS. It is assumed that $1 \leq U_c \leq T_c$ so that the precoding can be used to control the interference caused among the CUEs \cite{Bjornson-2014-a-SPM}.

In addition to the CUEs, there are other single-antenna users that bypass the BS and communicate pairwise with each other using a D2D communication mode. The locations of the D2D transmitters (D2D~Tx) are modeled by a homogeneous PPP $\Phi$ with density $\lambda_d$ in $\mathbb{R}^2$.\footnote{The assumption that the D2D~Tx are distributed in the whole $\mathbb{R}^2$ plane removes any concern about the boundary effects and makes the model more mathematically tractable. The boundary effects are local effects in which users at the network boundary experience less interference than the ones closer to the center, because they have fewer neighbors.} This means that the average number of D2D~Tx per unit area is $\lambda_d$ and these users are uniformly distributed in that area. The D2D receiver (D2D~Rx) is randomly located in an isotropic direction with a fixed distance away from its corresponding D2D~Tx---a model that is similar to the one considered in \cite{Lee-2014-JSAC}. The system setup is illustrated in Fig.~\ref{figure:sysMod_massiveMIMO_D2D}.

Let $R_{k,j}$ denote the distance between the $j$-th D2D~Tx to the $k$-th D2D~Rx. %
The performance analysis for D2D users is carried out for a typical D2D user, which is denoted by the index  $0$. The typical D2D user is an arbitrary D2D user located in the cell and its corresponding receiver is positioned in the origin. The results for a typical user show the statistical average performance of the network \cite{Haenggi-Stochastic}. Therefore, for any performance metric derivation, the D2D users inside the cell are considered and the ones outside the cell are only taken into account as sources of interference. Note that we neglect potential interference from other BSs and leave the multi-cell case for future work. This is because the interference from D2D transmissions is likely to be much stronger than the interference from other BSs. We assume equal power allocation for both CUEs and D2D users. Let $P_c$ denote the total transmit power of the BS, then the transmit power per CUE is $\frac{P_c}{U_c}$. %
The transmit power of the D2D~Tx is denoted by $P_d$.

Let $\vect{h}_j \in \mathbb{C}^{T_c \times 1}$ be the normalized channel response between the BS and the $j$-th CUE, for $j \in \{0, \dots, U_c-1 \}$. These channels are modeled as Rayleigh fading such that $\vect{h}_j \sim\mathcal{CN}(\mathbf{0},\mathbf{I})$, where $\mathcal{CN}(\cdot,\cdot)$ denotes a circularly symmetric complex Gaussian distribution. Perfect instantaneous channel state information (CSI) is assumed in this work for analytic tractability, but imperfect CSI is a relevant extension.  Linear downlink precoding is considered at the BS based on the zero-forcing (ZF) scheme that cancels the interference between the CUEs \cite{Bjornson-2014-a-SPM}. The precoding matrix is denoted by $\mathbf{V}=[\vect{v}_0, \dots, \vect{v}_{Uc-1}]\in \mathbb{C}^{T_c \times U_c}$ in which each column $\vect{v}_j$ is the normalized transmit precoding vector assigned to the CUE $j$. Let $\vect{f}_{0,\textrm{BS}} \in \mathbb{C}^{T_c \times 1}$ be the channel response from the BS to D2D~Rx and let it be Rayleigh fading as  $\vect{f}_{0,\textrm{BS}} \sim \mathcal{CN}(\mathbf{0},\mathbf{I})$.
Moreover, let $r_j \in \mathbb{C}$ and $\vect{s} \in \mathbb{C}^{U_c \times 1}$ denote the transmitted data signals intended for a D2D~Rx and the CUEs, respectively. Since each user requests different data, the transmitted signals can be modeled as zero-mean and uncorrelated with $\mathbb{E}\big[|r_j|^2\big]=P_d$ and $\mathbb{E}\big[||\vect{s}||^2\big]=P_c$.
The fading channel response between the $j$-th D2D~Tx and the $k$-th D2D~Rx is denoted by $g_{k,j} \in \mathbb{C}$ where $g_{k,j} \sim\mathcal{CN}(0,1)$. Moreover, $R_{0,\textrm{BS}}$ denotes the random distance between the typical D2D~Rx and the BS. The pathloss is modeled as $A_i d^{-\alpha_i}$ with $i \in \{c,d\}$, where index $c$ indicates the pathloss between a user and the BS and index $d$ gives the pathloss between any two users. $A_i$ and $\alpha_i$  are the pathloss coefficient and exponent, respectively, where we assume $\alpha_i>2$. %
The received signal at the typical D2D~Rx is
\begin{align}
y_{d,0} &= \sqrt{A_d } R_{0,0}^{-\alpha_d/2} g_{0,0} r_0 +\underbrace{ \sqrt{A_c } R_{0,\textrm{BS}}^{-\alpha_c/2} \vect{f}_{0,\textrm{BS}}^H \mathbf{V} \vect{s} }_{\textrm{Interference from the BS}} + \underbrace{\sqrt{A_d } \sum_{j\neq 0} R_{0,j}^{-\alpha_d/2} g_{0,j} r_j}_{\textrm{Interference from other D2D users}} + \eta_d,
\end{align}
where $\eta_d$ is zero-mean additive white Gaussian noise with  power  $N_0 = \tilde{N}_0 B_w$, $\tilde{N}_0$ is the power spectral density of the white Gaussian noise, and $B_w$ is the channel bandwidth. %
For given channel realizations, the signal-to-interference-plus-noise ratio (SINR) at the typical D2D~Rx is
\begin{align}\label{eq:SINR_d2d}
\mathrm{SINR}_d = \frac{P_d  R_{0,0}^{-\alpha_d}|g_{0,0}|^2}{I_{\textrm{BS},0} + I_{d,0} +  \frac{N_0}{A_d}},
\end{align}
in which both the numerator and the denominator have been normalized by $A_d$. $I_{\textrm{BS},0}$ is the received interference power from the BS and $I_{d,0}$ is the received interference power from other D2D users that transmit simultaneously which are defined as
\begin{align}
I_{\textrm{BS},0} &\triangleq \frac{\zeta R_{0,\textrm{BS}}^{-\alpha_c}}{A_d}\|\vect{f}_{0,\textrm{BS}}^H \mathbf{V}\|^2, \label{eq:I_BS0} \\
I_{d,0} &\triangleq \sum_{j\neq 0} P_d R_{0,j}^{-\alpha_d} |g_{0,j}|^2, \label{eq:I_d0}
\end{align}
where
\begin{equation}
\zeta \triangleq A_c \frac{P_c}{U_c}. \label{eq:zeta-def}
\end{equation}

Let $D_{0,k}$ and $e_{0,k}\in \mathbb{C}$ with $e_{0,k}\sim\mathcal{CN}(0,1)$  be the distance and fading channel response between a typical CUE and the $k$-th D2D~Tx, respectively, and let $D_{0,\textrm{BS}}$ denote the distance between a typical CUE and the BS. Then, the received signal at the typical CUE is
\begin{align}
y_{c,0} &= \sqrt{A_c} D_{0,\textrm{BS}}^{-\alpha_c/2} \vect{h}_{0}^H \mathbf{V} \vect{s}  + \underbrace{\sqrt{A_d } \sum_{j} D_{0,j}^{-\alpha_d/2} e_{0,j} r_j}_{\textrm{Interference from all D2D users}} + \eta_c,
\end{align}
where $\eta_c$  is zero-mean additive white Gaussian noise with power $N_0$. Then, the corresponding SINR for the typical CUE is
\begin{align}
\mathrm{SINR}_c = \frac{|\vect{h}_{0}^H \vect{v}_{0}|^2}{\frac{A_d }{\zeta} D_{0,\textrm{BS}}^{\alpha_c}(I_{d,c} + \frac{N_0}{A_d })}, \label{eq:SINR_CUE}
\end{align}
where
\begin{equation}\label{eq:inf_dc}
I_{d,c} \triangleq \sum_{j} P_d D_{0,j}^{-\alpha_d} |e_{0,j}|^2
\end{equation}
is the received interference power from all D2D users (normalized by $A_d$).

\section{Performance Analysis}
\label{sec:performance_analysis}
In this section, we first introduce the performance metrics that are considered in this paper. Then we proceed to derive the coverage probability  for both CUEs and D2D users which are needed to compute these metrics.

\subsection{Performance Metrics}
\label{sec:metrics}

In this paper, two main performance metrics for the network are considered:  the average sum rate (ASR) and energy efficiency (EE). %
The ASR is obtained from total rates of both D2D users and CUEs  as
\begin{equation}
\text{ASR}= U_c \bar{R}_c +  \pi R^2 \lambda_d \bar{R}_d,
\label{eq:ASR}
\end{equation}
where $\pi R^2 \lambda_d$ is the average number of D2D users in the cell and $\bar{R}_t$ with $t\in\{c,d\}$ denotes the average rates of the CUEs and D2D users, respectively. $\bar{R}_t$ for both cellular and D2D users is computed as the successful transmission rate by
\begin{equation}\label{eq:Avg_Rate1}
\bar{R}_t = \underset{\beta_t \geq 0}{\mathrm{sup}}~B_w \log_{2}(1 + \beta_t) \mathrm{P}^t_{\mathrm{cov}} (\beta_t)
\end{equation}
where
\begin{equation}\label{eq:PsuccDef}
\mathrm{P}^t_{\mathrm{cov}} (\beta_t) = \mathrm{Pr}\big\{\mathrm{SINR}_{t} \geq \beta_t \big\}
\end{equation}
is the coverage probability  when the received SINR is higher than a specified threshold $\beta_t$ needed for successful reception. Note that $\mathrm{SINR}_{t}$ contains random channel fading and random user locations. Finding the supremum guarantees the best constant rate for the D2D users and the CUEs.  If we know the coverage probability  ($\mathrm{P}^t_{\mathrm{cov}} (\beta_t)$), \eqref{eq:Avg_Rate1} can easily be computed by using line search for each user type independently. Moreover, \eqref{eq:Avg_Rate1} is easily achievable in practice since the modulation and coding is performed without requiring that every transmitter knows the interference characteristics at its receiver.

Energy efficiency is defined as the benefit-cost ratio between the ASR and the total consumed power:
\begin{equation}
\text{EE} = \frac{\text{ASR}}{\text{Total power}}. \label{eq:EE}
\end{equation}
For the total power consumption, we consider a detailed  model described in \cite{Bjornson-2014-arxiv}:
\begin{align}
\text{Total power}&=\frac{1}{\eta} \big(P_c +\lambda_d \pi R^2 P_d\big) + C_0  + T_c C_1 + \big(U_c + 2 \lambda_d \pi R^2\big)C_2,\label{eq:P_tot}
\end{align}
where $P_c +\lambda_d \pi R^2 P_d$ is the total transmission power averaged over the number of D2D users, $\eta$ is the amplifier efficiency ($0 < \eta \leq 1$), $C_0$ is the load independent power consumption at the BS, $C_1$ is the power consumption per BS antenna, $C_2$ is the power consumption per user device, and $U_c + 2 \lambda_d \pi R^2$ is the average number of active users.

In order to calculate the ASR and EE, we need to derive the coverage probability  for both cellular and D2D users. The analytic derivation of these expressions is one of the main contributions of this paper.

\subsection{Coverage Probability of D2D Users}

We first derive the expression for the coverage probability  of D2D users.

\begin{proposition} \label{proposition:P_succ_d2d}
The approximate coverage probability  for a typical D2D user is given by
\begin{align}
\mathrm{P}^d_{\mathrm{cov}}(\beta_d) &=    \frac{ (\kappa\beta_d)^{2/\alpha_c}}{R^2}\left(y^{U_c + \frac{2}{\alpha_c}-1} (1-y)^{- \frac{2}{\alpha_c}}  - \Big(U_c + \frac{2}{\alpha_c}-1\Big) \mathcal{B}\Big(y; U_c + \frac{2}{\alpha_c}-1, 1-\frac{2}{\alpha_c}\Big) \right) \notag \\
&\;\quad \cdot \exp\bigg(- \frac{\pi \lambda_d R_{0,0}^2}{\mathrm{sinc}(\frac{2}{\alpha_d})} \beta_d^{2/\alpha_d}\bigg) \exp\bigg(-\frac{\beta_d} {\bar{\gamma}_d}\bigg),
\label{eq:P_succ_d2d}
\end{align}
where $\kappa \triangleq \frac{\zeta}{P_d A_d R_{0,0}^{-\alpha_d}}$ with $\zeta$ defined in \eqref{eq:zeta-def}, $y \triangleq \frac{1}{\kappa \beta_d R^{-\alpha_c} + 1}$, $\mathrm{sinc}(x) = \frac{\sin(\pi x)}{\pi x}$, $\bar{\gamma}_d = \frac{A_d  R_{0,0}^{-\alpha_d}P_d}{N_0}$ is the average D2D SNR, and $\mathcal{B}(x;a,b)$ is the incomplete Beta function.
\end{proposition}
\begin{IEEEproof}
The proof is given in Appendix~\ref{sec:proof_Pcov_d2d}.
\end{IEEEproof}

The coverage probability  expression in Proposition~\ref{proposition:P_succ_d2d} allows us to compute the average data rate of a typical D2D user in \eqref{eq:Avg_Rate1}. We note that \eqref{eq:P_succ_d2d} is actually a tight approximation and its tightness is evaluated in Section~\ref{sec:results}. %
From the expression in \eqref{eq:P_succ_d2d}, we make several observations as listed below.

\begin{remark} \label{remark:P_succ_d2d_highSNR}
In the high-SNR regime for the D2D users where $\bar{\gamma}_d  \gg \beta_d$, the last term in \eqref{eq:P_succ_d2d} converges to one, i.e., $\exp\left(- \frac{\beta_d} {\bar{\gamma}_d}\right) \to 1$, and we have
\begin{align}
\mathrm{P}^d_{\mathrm{cov}}(\beta_d) &=    \frac{ (\kappa\beta_d)^{2/\alpha_c}}{R^2}\left(y^{U_c + \frac{2}{\alpha_c}-1} (1-y)^{- \frac{2}{\alpha_c}}  - \Big(U_c + \frac{2}{\alpha_c}-1\Big) \mathcal{B}\Big(y; U_c + \frac{2}{\alpha_c}-1, 1-\frac{2}{\alpha_c}\Big) \right) \notag \\
&\;\quad \cdot \exp\bigg(- \frac{\pi \lambda_d R_{0,0}^2}{\mathrm{sinc}(\frac{2}{\alpha_d})} \beta_d^{2/\alpha_d}\bigg).
\label{eq:P_succ_d2d_highSNR}
\end{align}
This can also be referred to as the interference-limited regime.
\end{remark}

\begin{remark} \label{remark:P_succ_d2d_lambda_d}
The coverage probability  of a typical D2D user is a decreasing function of the D2D density $\lambda_d$. Because higher $\lambda_d$ results in more interference among D2D users. In particular, it can be seen that $\mathrm{P}^d_{\mathrm{cov}}$ in \eqref{eq:P_succ_d2d} is a function of $\lambda_d$ through $\exp(-C\lambda_d)$ with $C\triangleq\frac{\pi R_{0,0}^2 \beta_d^{2/\alpha_d}}{\mathrm{sinc}(\frac{2}{\alpha_d})} > 0$. Thus, if $\lambda_d \to \infty$, $\mathrm{P}^d_{\mathrm{cov}} \to 0$.

Recall that in our model, the D2D~Rx is associated to the D2D~Tx which is located at a fixed distance away. However, if we had assumed that the D2D~Rx's association to a D2D~Tx is based on, for example, the shortest distance or the maximum SINR, then the $\mathrm{P}^d_{\mathrm{cov}}$ would have been unaffected by the D2D density (in the high-interference regime).
\end{remark}

Now, considering the number of BS antennas or the number of CUEs as variables, we have the following behavior of the D2D coverage probability.

\begin{remark} \label{remark:P_succ_d2d_Tc_Uc}
$\mathrm{P}^d_{\mathrm{cov}}$ is not affected by the number of BS antennas $T_c$. The BS antennas are used to cancel out the interference among CUEs and they do not have any impact on D2D users' performance as long as the number of CUEs $U_c$ is constant and does not vary with the number of BS antennas $T_c$. The coverage probability  of a typical D2D user $\mathrm{P}^d_{\mathrm{cov}}$ is a decreasing function of $U_c$. However, increasing the number of CUEs have a small effect on D2D users' performance. This is due to the fact that the resulting interference from the BS to D2D users does not change significantly by increasing the number of CUEs as the transmit power of the BS is the same irrespective of the number of users and the precoding is independent of the D2D channels. Thus, a change of $U_c$ will only change the distribution of the interference but not its average.
\end{remark}

Next we comment on how changes in the transmit powers of the BS and D2D~Tx as well as the distance between D2D user pairs affect the coverage probability  of D2D users.

\begin{remark} \label{remark:P_succ_d2d_PcPd}
$\mathrm{P}^d_{\mathrm{cov}}$ is a decreasing function of the ratio between the transmit power of the BS and of the D2D users, i.e., $\frac{P_c}{P_d}$, which is part of the first term in \eqref{eq:P_succ_d2d} and corresponds to the interference from the BS. For instance, if we fix $P_c$ and decrease $P_d$, the coverage probability  for D2D users decreases as the interference from the BS would be the dominating factor. At the same time, if we decrease $P_c$, it would improve the coverage of D2D users.
\end{remark}

\begin{remark} \label{remark:P_succ_d2d_Uc}
$\mathrm{P}^d_{\mathrm{cov}}$ is a decreasing function of the distance between D2D~Tx-Rx pairs $R_{0,0}$ and the cell radius $R$. Increasing the cell radius with the same D2D user density reduces the effect of the interference from the BS. Also by decreasing the distance between D2D~Tx-Rx pairs, it is evident that a better performance for D2D users can be obtained.
\end{remark}

Using Proposition~\ref{proposition:P_succ_d2d}, the following corollary provides the optimal D2D user density that maximizes the D2D ASR, i.e., $\pi R^2 \lambda_d \bar{R}_d$, where $\bar{R}_d$ is given in \eqref{eq:Avg_Rate1}.

\begin{corollary} \label{corollary:ASR_d2d_lambda_d_max}
For a given SINR threshold $\beta_d$, the optimal density of D2D users $\lambda_d^*$ that maximizes the D2D ASR is
\begin{align}
\lambda_d^*(\beta_d) = \frac{\mathrm{sinc}(\frac{2}{\alpha_d})}{\pi R_{0,0}^2}\beta_d^{-2/\alpha_d}. \label{eq:lambda_max}
\end{align}
\end{corollary}
\begin{IEEEproof}
Given the SINR threshold $\beta_d$ and using \eqref{eq:ASR}--\eqref{eq:Avg_Rate1}, the D2D ASR is
\begin{equation}
\pi R^2 \lambda_d B_w\log_2 (1+\beta_d) \mathrm{P}^d_{\mathrm{cov}}(\beta_d), \label{eq:D2D_ASR_optLambda_d}
\end{equation}
where $\mathrm{P}^d_{\mathrm{cov}}(\beta_d)$ is given in \eqref{eq:P_succ_d2d} and depends on $\lambda_d$ through an exponential function. Taking the derivative of \eqref{eq:D2D_ASR_optLambda_d} with respect to $\lambda_d$ and setting it to zero yields the optimal D2D user density $\lambda_d^*(\beta_d)$ given in \eqref{eq:lambda_max} that maximizes the D2D ASR.
\end{IEEEproof}

\subsection{Coverage Probability of Cellular Users}

Next, we compute the coverage probability  for CUEs.
\begin{proposition} \label{proposition:P_succ_cue}
The coverage probability  for a typical cellular user is given by
\begin{align}
\mathrm{P}^c_{\mathrm{cov}}(\beta_c) &= \mathbb{E}_{D_{0,\textrm{BS}}}\left[e^{-\frac{N_0}{A_d}s} \sum_{k=0}^{T_c - U_c} \frac{s^k}{k!} \sum_{i=0}^{k} \binom{k}{i} \left(\frac{N_0}{A_d }\right)^{k-i}(-1)^i
\;\Upsilon(\lambda_d,s,i)\right],
\label{eq:P_succ_cue}
\end{align}
with
\begin{align}
\Upsilon(\lambda_d,s,i) &= \exp\left(-C_d \lambda_d s^{2/\alpha_d} \right)  \sum_{(j_1,\ldots,j_i)\in\mathcal{J}} i! \prod_{\ell=1}^i\frac{1}{j_\ell!(\ell!)^{j_\ell}}\left(-C_d \lambda_d s^{\frac{2}{\alpha_d}-\ell}\prod_{q=0}^{\ell-1}\Big(\frac{2}{\alpha_d}-q\Big)\right)^{j_\ell}, \label{eq:Upsilon}
\end{align}
where $s\triangleq \frac{A_d}{\zeta} D_{0,\textrm{BS}}^{\alpha_c}\beta_c$ with $\zeta$ defined in \eqref{eq:zeta-def}, $C_d \triangleq \frac{\pi P_d^{2/\alpha_d}}{\mathrm{sinc}(\frac{2}{\alpha_d})}$, and
\begin{equation*}
\mathcal{J}\triangleq \bigg\{(j_1,\ldots,j_i): j_\ell \in \mathbb{Z}_{\geq 0}, \; \sum_{\ell=1}^i  \ell j_\ell = i \bigg\}.
\end{equation*}
\end{proposition}
\begin{IEEEproof}
The proof is given in Appendix~\ref{sec:proof_Pcov_cue}.
\end{IEEEproof}

This proposition gives an expression for the coverage probability  of CUEs in which there is only one random variable left. The expectation in  \eqref{eq:P_succ_cue} with respect to $D_{0,\textrm{BS}}$ is intractable to derive analytically but can be computed numerically. %
The analytical results of Proposition~\ref{proposition:P_succ_d2d} and Proposition~\ref{proposition:P_succ_cue} have been verified by Monte-Carlo simulations in Section~\ref{sec:results}. A main benefit of the analytic expressions (as compared to pure Monte-Carlo simulations with respect to all sources of randomness) is that they can be computed much more efficiently, which basically is a prerequisite for the multi-variable system analysis carried out in Section~\ref{sec:results}.

Next, we present some observations from the result in Proposition~\ref{proposition:P_succ_cue} as follows.
\begin{remark} \label{remark:P_succ_cue_inflim}
In the interference-limited regime where where $I_{d,c}\gg N_0$, the coverage probability  in \eqref{eq:P_succ_cue} for a typical cellular user is simplified to
\begin{align}
\mathrm{P}^c_{\mathrm{cov}}(\beta_c) &= \mathbb{E}_{D_{0,\textrm{BS}}}\left[\sum_{k=0}^{T_c - U_c} \frac{(-s)^k}{k!}\Upsilon(\lambda_d,s,k)\right].
\label{eq:P_succ_cue_inflim}
\end{align}
\end{remark}
The result obtained in Remark~\ref{remark:P_succ_cue_inflim} has a lower computational complexity compared to the expression in Proposition~\ref{proposition:P_succ_cue} and at the same time it is a tight approximation for Proposition~\ref{proposition:P_succ_cue}. This can be observed from the denominator of the \eqref{eq:SINR_CUE} where the term $\frac{N_0}{A_d }\approx 0$.

\begin{remark} \label{remark:P_succ_cue_lambda_dinf}
The coverage probability  of a typical CUE $\mathrm{P}^c_{\mathrm{cov}}(\beta_c)$  is a decreasing function of the D2D user density $\lambda_d$. From Proposition~\ref{proposition:P_succ_cue}, only $\Upsilon(\lambda_d,s,i)$ is a function of $\lambda_d$ which is composed of an exponential term in $\lambda_d$ multiplied by a polynomial term in $\lambda_d$.  Thus, if $\lambda_d \to \infty$, the exponential term which has a negative growth dominates the polynomial term and  $\mathrm{P}^c_{\mathrm{cov}}(\beta_c) \to 0$.
\end{remark}

We proceed to analyze the behavior of Proposition~\ref{proposition:P_succ_cue} by considering a number of special cases.

\begin{corollary} \label{corollary:col_P_succ_cue_TcequalUc}
If $T_c = U_c$, the coverage probability  for a typical cellular user is given by
\begin{align}
\mathrm{P}^c_{\mathrm{cov}}(\beta_c) &= \mathbb{E}_{D_{0,\textrm{BS}}}\left[\exp\bigg(-\frac{N_0}{A_d}s - C_d \lambda_d s^{2/\alpha_d}\bigg)\right], \label{eq:P_succ_cue_TcequalUc}
\end{align}
where $s = \frac{A_d}{\zeta} D_{0,\textrm{BS}}^{\alpha_c} \beta_c$ and $C_d=\frac{\pi P_d^{2/\alpha_d}}{\mathrm{sinc}(\frac{2}{\alpha_d})}$.
\end{corollary}
\begin{IEEEproof}
\eqref{eq:P_succ_cue_TcequalUc} follows directly from \eqref{eq:P_succ_cue} by setting $T_c-U_c = 0$.
\end{IEEEproof}

\begin{corollary} \label{corollary:P_succ_cue_TcUcinf}
If $(T_c - U_c) \to \infty$, the coverage probability  for a typical cellular user tends to one, that is,
\begin{align}
\lim_{(T_c - U_c) \to \infty}~\mathrm{P}^c_{\mathrm{cov}}(\beta_c) = 1.
\label{eq:P_succ_cue_TcUcinf}
\end{align}
\end{corollary}
\begin{IEEEproof}
Let $m=T_c - U_c$. Substituting $\mathrm{SINR}_c$ from \eqref{eq:SINR_CUE}  into \eqref{eq:PsuccDef}, we have
\begin{align*}
\lim_{m \to \infty}~\mathrm{P}^c_{\mathrm{cov}}(\beta_c) &= \lim_{m \to \infty}~\mathrm{Pr}\left\{ |\vect{h}_{0}^H \vect{v}_{0}|^2 \geq \frac{A_d}{\zeta} D_{0,\textrm{BS}}^{\alpha_c}\Big(I_{d,c} + \frac{N_0}{A_d}\Big)\beta_c \right\} \\
&\eqtop{a} \lim_{m \to \infty}~\mathbb{E}_{D_{0,\textrm{BS}}, I_{d,c}}\left[e^{-\frac{A_d}{\zeta} D_{0,\textrm{BS}}^{\alpha_c}(I_{d,c} + \frac{N_0}{A_d})\beta_c }  \sum_{k=0}^{m} \frac{1}{k!}\bigg(\frac{A_d}{\zeta} D_{0,\textrm{BS}}^{\alpha_c}\Big(I_{d,c} + \frac{N_0}{A_d}\Big)\beta_c\bigg)^k\right]  \\
&\eqtop{b} \lim_{m \to \infty}~\mathbb{E}_{D_{0,\textrm{BS}}, I_{d,c}}\left[e^{-z} \sum_{k=0}^{m} \frac{z^k}{k!}\right] \\
&\eqtop{c} \mathbb{E}_{D_{0,\textrm{BS}}, I_{d,c}}\left[\lim_{m \to \infty} e^{-z} \sum_{k=0}^{m} \frac{z^k}{k!}\right] \\
&\eqtop{d} \mathbb{E}_{D_{0,\textrm{BS}}, I_{d,c}}\left[e^{-z} e^{z} \right] = 1,
\end{align*}
where $(a)$ follows from the CCDF of $|\vect{h}_{0}^H \vect{v}_{0}|^2$ with $2|\vect{h}_{0}^H \vect{v}_{0}|^2 \sim \chi^2_{2}$ given $D_{0,\textrm{BS}}$ and $I_{d,c}$. Step $(b)$ follows from setting $z = \frac{A_d}{\zeta} D_{0,\textrm{BS}}^{\alpha_c}(I_{d,c} + \frac{N_0}{A_d})\beta_c$.  Step $(c)$ is obtained from the dominated convergence theorem which allows for an interchange of limit and expectation and step $(d)$ is due to the fact that $\sum_{k=0}^{\infty} \frac{z^k}{k!} = e^z$.
\end{IEEEproof}

In the results so far, we have discussed the case where there exist some D2D users as underlay to the cellular network, that is, $\lambda_d \neq 0$, However, it is interesting to see what can be achieved without D2D users.

\begin{corollary} \label{corollary:col_P_succ_cue_lambda_dzero}
If $\lambda_d= 0$, the coverage probability  for a typical cellular user is given by
\begin{align}
\mathrm{P}^c_{\mathrm{cov}}(\beta_c) &= \frac{2}{\alpha_c R^2}\Gamma\left(\frac{2}{\alpha_c}\right)\left(\frac{N_0}{\zeta} \beta_c \right)^{-2/\alpha_c}
 \sum_{k=0}^{T_c-U_c}\binom{\frac{2}{\alpha_c}+k-1}{k}, \label{eq:P_succ_cue_lambda_dzero}
\end{align}
where $\Gamma(\cdot)$ is the Gamma function and $\zeta$ is defined in \eqref{eq:zeta-def}.
\end{corollary}
\begin{IEEEproof}
Substituting $\mathrm{SINR}_c$ from \eqref{eq:SINR_CUE} into \eqref{eq:PsuccDef} and setting $\lambda_d= 0$, we have
\begin{align}
\mathrm{P}^c_{\mathrm{cov}}(\beta_c) &= \mathrm{Pr}\left\{ |\vect{h}_{0}^H \vect{v}_{0}|^2 \geq \frac{D_{0,\textrm{BS}}^{\alpha_c}}{\zeta} N_0 \beta_c \right\} \nonumber\\
&\eqtop{a} \mathbb{E}_{z}\Bigg[\sum_{k=0}^{T_c-U_c}\frac{l^k}{k!}z^k e^{-lz} \Bigg] \nonumber\\
&\eqtop{b}\frac{2}{\alpha_c R^2}\Gamma\left(\frac{2}{\alpha_c}\right)
 \sum_{k=0}^{T_c-U_c}\frac{(-l)^k}{k!} ~\frac{\mathrm{d}^k}{{\mathrm{d}l}^k} ~l^{-2/\alpha_c},
\end{align}
where $(a)$ follows from the CCDF of $|\vect{h}_{0}^H \vect{v}_{0}|^2$ with $2|\vect{h}_{0}^H \vect{v}_{0}|^2 \sim \chi^2_{2}$ given $D_{0,\textrm{BS}}$ and setting $l= \frac{N_0}{\zeta} \beta_c$ and $z = D_{0,\textrm{BS}}^{\alpha_c}$ with PDF $f(z)=\frac{2}{\alpha_cR^2}z^{\frac{2}{\alpha_c}-1}$. Step $(b)$ follows from taking the expectation with respect to $z$ which is similar to the expression in \eqref{eq:E_Idc} with the Laplace transform $ \mathcal{L}_{z}(l) = \frac{2}{\alpha_c R^2}\Gamma\big(\frac{2}{\alpha_c}\big)l^{-2/\alpha_c}$. Simplifying the $k$-th derivative to $\frac{\mathrm{d}^k}{\mathrm{d}l^k} ~l^{-2/\alpha_c} = (-1)^k l^{-\frac{2}{\alpha_c}-k} \prod_{i=0}^{k-1}\big(\frac{2}{\alpha_c}+i\big)$ and using the identity $\frac{1}{k!}\prod_{i=0}^{k-1}\big(\frac{2}{\alpha_c}+i\big) = \binom{\frac{2}{\alpha_c}+k-1}{k}$, \eqref{eq:P_succ_cue_lambda_dzero} follows.
\end{IEEEproof}

The closed-form results in Corollary~\ref{corollary:col_P_succ_cue_lambda_dzero} for $\lambda_d= 0$ depends only on noise rather than interference and perhaps can result in higher ASR for CUEs. The ASR for $\lambda_d>0$ also depends on noise but its impact is much smaller.   However, we note that this result is obtained for a single cell scenario. Thus, comparing Proposition~\ref{proposition:P_succ_cue} and Corollary~\ref{corollary:col_P_succ_cue_lambda_dzero} and evaluating the potential performance gain/loss due to introducing D2D communications would make more sense in a multi-cell scenario.

Using the results from Proposition~\ref{proposition:P_succ_d2d} and Proposition~\ref{proposition:P_succ_cue}, we proceed to evaluate the network performance in terms of the ASR and EE from \eqref{eq:ASR} and \eqref{eq:EE}, respectively.

\section{Numerical Results}
\label{sec:results}

\begin{table}[tp]
\small
\centering
\caption{System and simulation parameters.}
\label{table:Sim_param}
\begin{tabular}{@{} l c c  @{}} \toprule
  Description & Parameter & Value\\ \midrule
  D2D TX power & $P_d$ & $6$ dBm\\
  BS TX power &  $P_c$ & $30$ dBm\\
  Cell radius & $R$ & $500$ m\\
  Bandwidth & $B_w$  & $20$ MHz\\
  Thermal noise power & $N_0$ & $-131$ dBm\\
  Noise figure in  UE & $F$ & $5$ dB\\
  Carrier frequency & $f_c$  & $2$ GHz\\
  D2D pair distance & $R_{0,0}$ & $35$ m\\
  Pathloss exponent betw.\ devices & $\alpha_d$  & $3$\\
  Pathloss exponent betw.\ BS--device & $\alpha_c$  & $3.67$\\
  Pathloss coefficient betw.\ devices & $A_d$  & $38.84$ dB\\
  Pathloss coefficient betw.\ BS--device & $A_c$  & $30.55$ dB\\
  Amplifier efficiency & $\eta$  & $0.3$\\
  Load-independent power in BS & $C_0$  & $5$ W\\
  Power per BS antenna & $C_1$  & $0.5$ W\\
  Power per UE handset & $C_2$  & $0.1$ W\\
  Monte-Carlo runs & MC  & $5000$\\\bottomrule
\end{tabular}
\vspace{-6mm}
\end{table}

In this section, we assess the performance of the setup in Fig.~\ref{figure:sysMod_massiveMIMO_D2D} in terms of ASR and EE using numerical evaluations. As we pointed out in \secref{sec:performance_analysis}, many parameters affect these performance metrics.
Initially, we consider the EE and the ASR as functions of
three key parameters, namely, the number of BS antennas $T_c$, the density of D2D users $\lambda_d$, and the number of cellular users $U_c$. We  show the individual effect of these system parameters on the two performance metrics while other parameters such as BS transmit power $P_c$,  D2D transmit power $P_d$, and distance between D2D~Tx-Rx pair $R_{0,0}$ are fixed. Later on, we also comment on the choice of these fixed parameters. The system and simulation parameters are given in Table~\ref{table:Sim_param}.

\begin{figure*}[tp]
\centering
\subfloat[]{
\includegraphics[width=.48\linewidth]{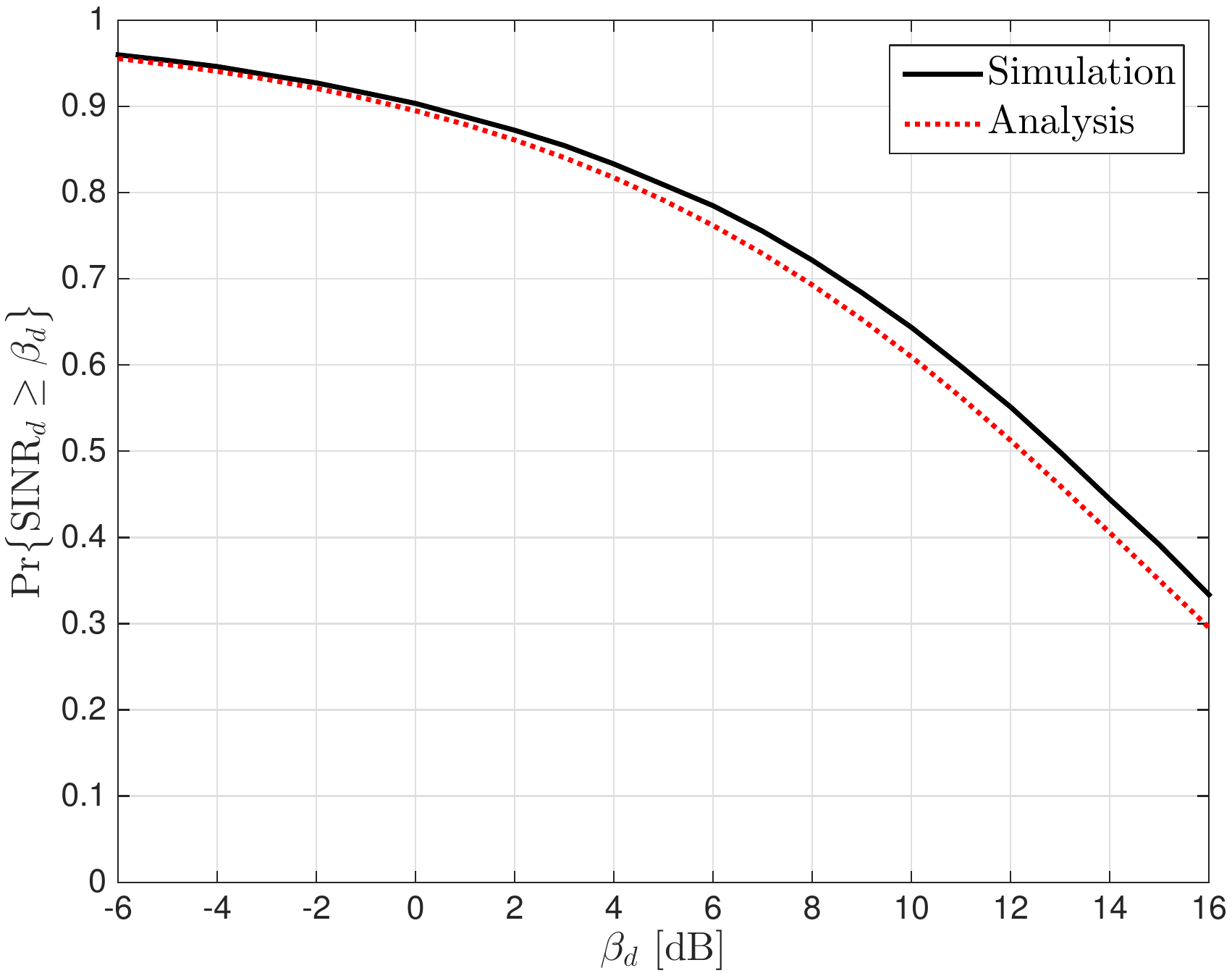}
\label{fig:P_cov_d2d}
}
\hfil
\subfloat[]{
\includegraphics[width=.48\linewidth]{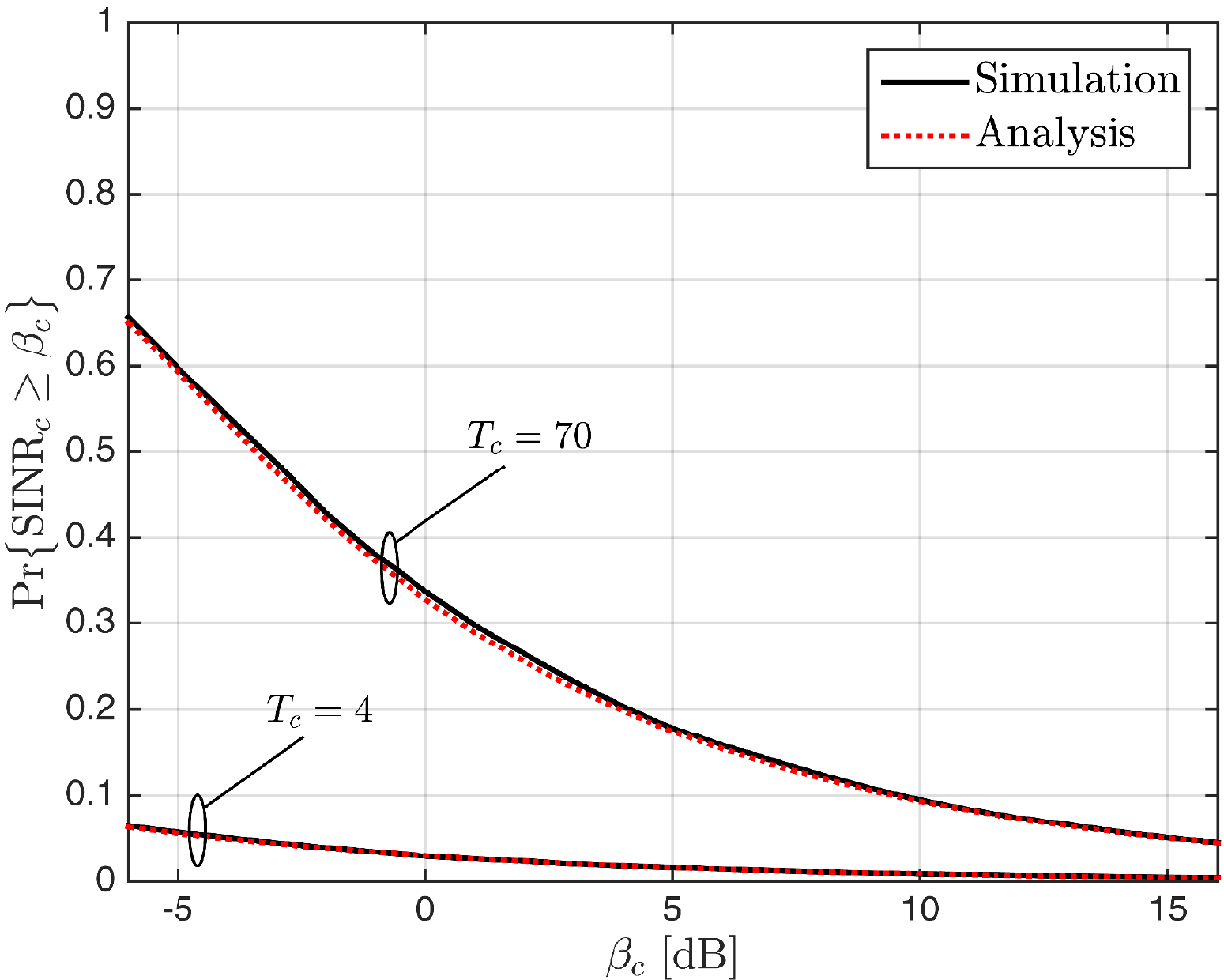}
\label{fig:P_cov_cue}
}
\caption{Coverage probability  as a function of $\beta_t$, $t\in\{d,c\}$: analysis versus Monte-Carlo simulations for (a) D2D users with $\lambda_d=10^{-5}$ and (b) CUEs with $\lambda_d=10^{-5}$ and $T_c\in\{4,70\}$.}
\label{fig:P_cov_analysis_sim}
\end{figure*}

Before we proceed to the performance evaluation, we verify the analytical results of Proposition~\ref{proposition:P_succ_d2d} and Proposition~\ref{proposition:P_succ_cue} by Monte-Carlo simulations. As depicted in \figref{fig:P_cov_analysis_sim}, simulation results closely follow the analytical derivations. The small gap in Fig.~\ref{fig:P_cov_d2d} is due to the spatial interference correlation resulting from the fact that multiple interfering streams are coming from the same location, hence, the Chi-squared distribution in \eqref{eq:normfHv} is an approximation. This is a quite standard approximation in analyzing MIMO systems \cite{Dhillon-2013-TWC}. Moreover, in the simulations, the locations of the D2D~Tx are generated in an area with radius $10R$ according to the PPP as opposed to our analytical assumption that they are located in the whole $\mathbb{R}^2$ region. This assumption reduces the interference as compared to our analytical results and thus improves the coverage probability as can be seen in Fig.~\ref{fig:P_cov_d2d}.

We consider two scenarios corresponding to the number of CUEs $U_c$ in our evaluations. First, we assume that $U_c$ is chosen as a function of the number of BS antennas $T_c$. Then, we move on to the case where we fix the number of CUEs and study the tradeoffs among other parameters. Both scenarios are relevant in the design of massive MIMO systems. In order to speed up the numerical computations, we neglected the terms that are very small.

\subsection{Number of CUEs as a Function of the Number of BS Antennas}

In this scenario, we assume that there is a fixed ratio between the number of CUEs  $U_c$ and the number of BS antennas $T_c$. We assume this ratio to be $\frac{T_c}{U_c}=5$. Simply put, to serve one additional user, we add five more antennas at the BS since the main gains from massive MIMO come from multiplexing of many users rather than only having many antennas.

\begin{figure}[tp]
\centering
\includegraphics[width=0.65\columnwidth]{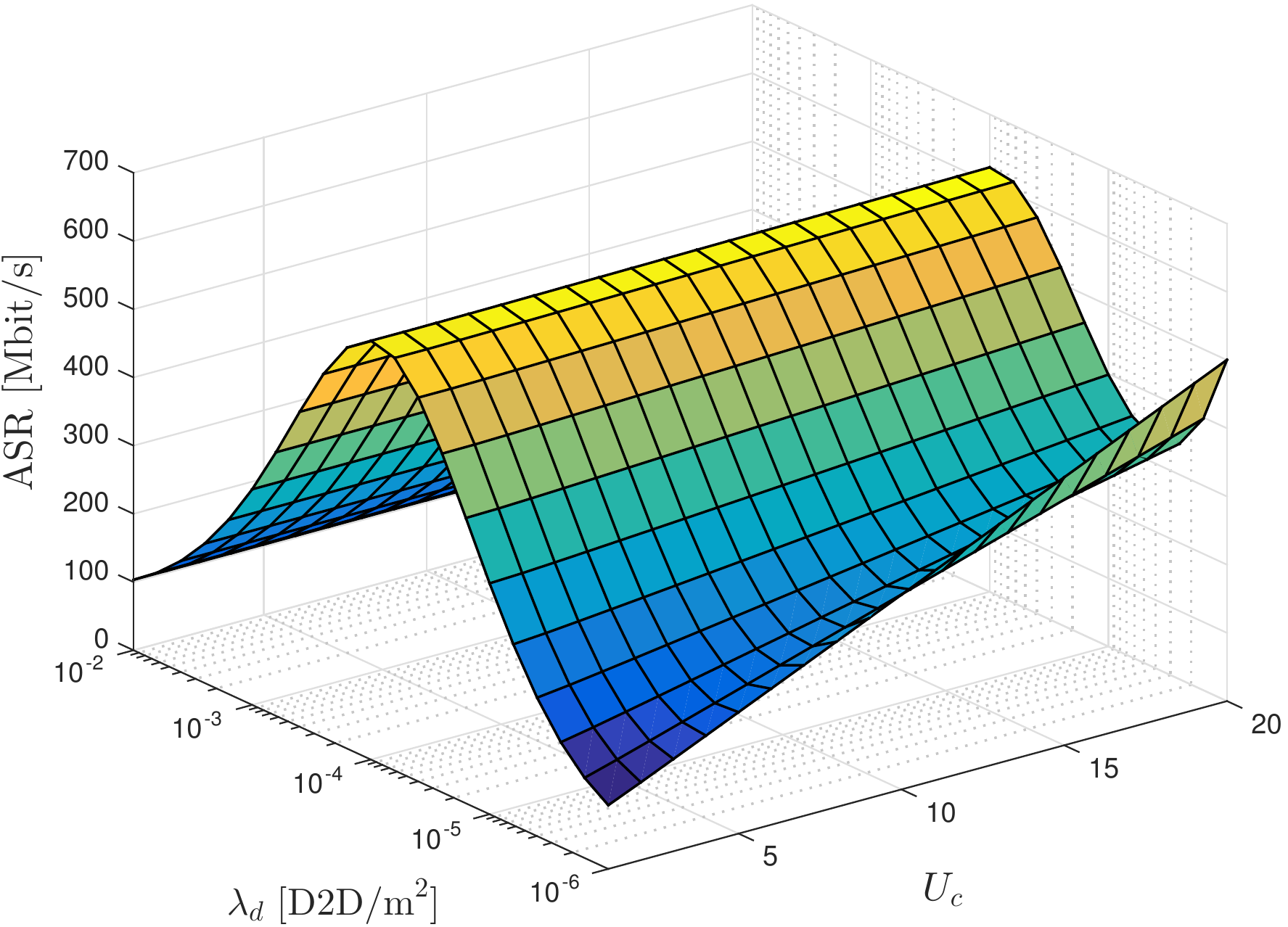}
\caption{ASR $\mathrm{[Mbit/s]}$ as a function of the number of CUEs $U_c$ and the D2D user density $\lambda_d$ for a fixed ratio $\frac{T_c}{U_c}= 5$.}
\label{fig:ASR_ld_uc_phi5}
\end{figure}

\begin{figure*}[tp]
\centering
\subfloat[]{
\includegraphics[width=.48\linewidth]{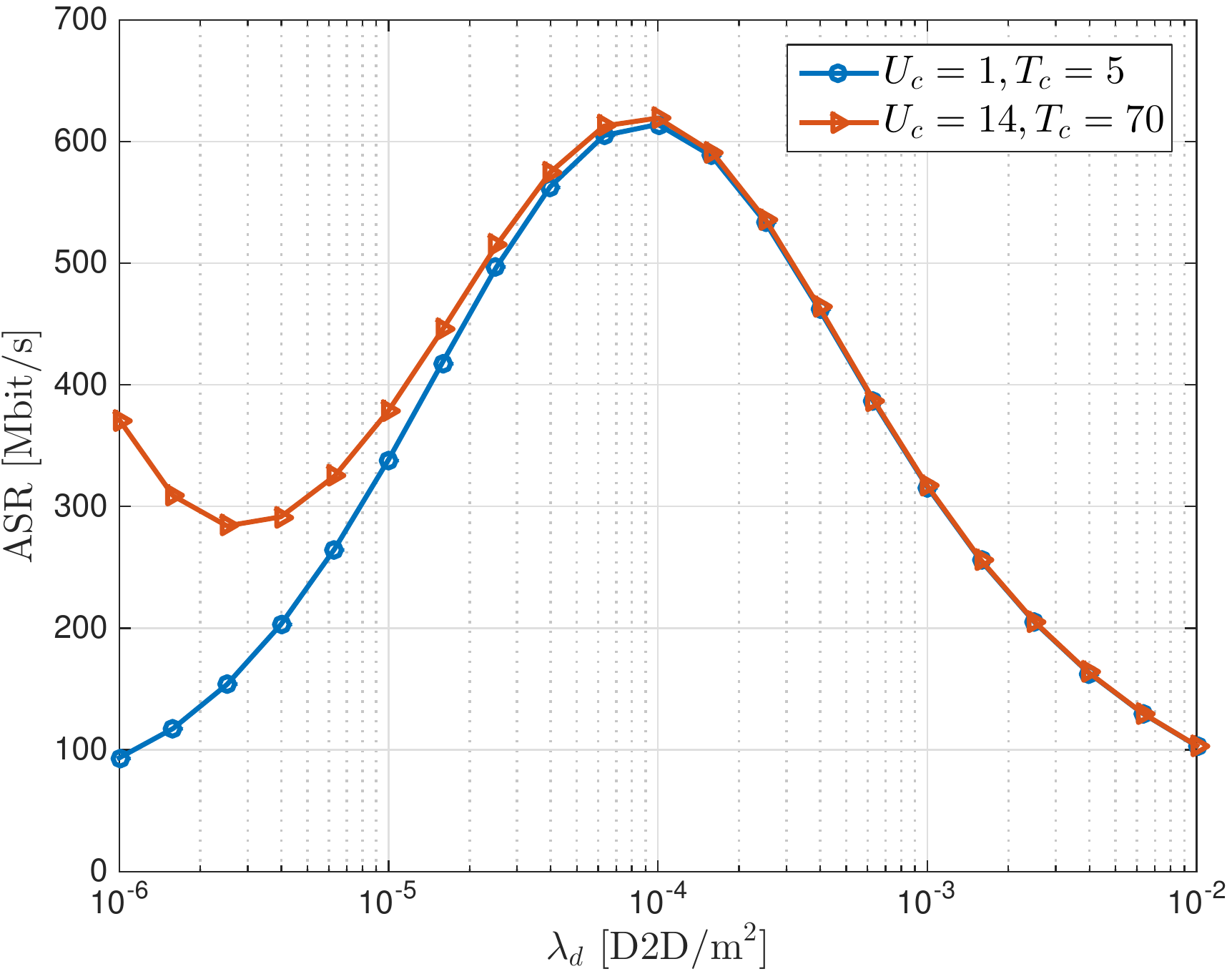}
\label{fig:ASR_ld_uc-1-14_phi5}
}
\hfil
\subfloat[]{
\includegraphics[width=.48\linewidth]{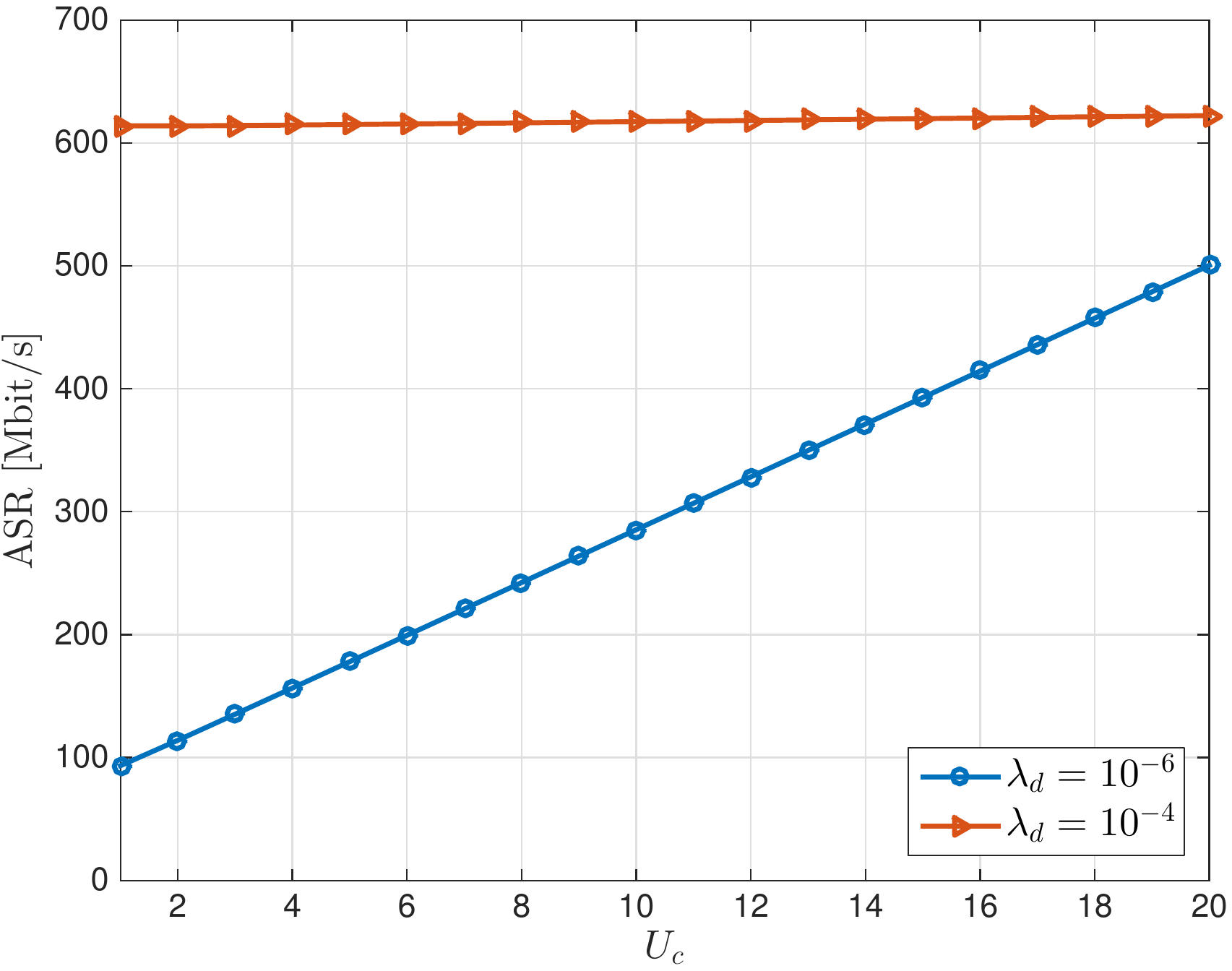}
\label{fig:ASR_ld-4-6_uc_phi5}
}
\caption{ASR $\mathrm{[Mbit/s]}$: (a) as a function of the D2D user density $\lambda_d$ for a fixed ratio $\frac{T_c}{U_c}= 5$ with the number of CUEs $U_c\in\{1,14\}$; (b) as a function of the number of CUEs $U_c$ with the D2D user density $\lambda_d\in\{10^{-6},10^{-4}\}$ for a fixed ratio  $\frac{T_c}{U_c}= 5$.}
\label{fig:ASR_2D}
\end{figure*}

\figref{fig:ASR_ld_uc_phi5}  shows the ASR as a function of the density of D2D users $\lambda_d$ and the number of CUEs  $U_c$, which is scaled by $T_c$.  It is observed that increasing  $U_c$, or equivalently $T_c$, always increases the ASR. In contrast, there is an optimal value of $\lambda_d$ as derived in Corollary~\ref{corollary:ASR_d2d_lambda_d_max} which results in the maximum ASR for all values of $U_c$ and appears approximately at $\lambda_d = 10^{-4}$. However, there is a difference in the shape of the ASR between the lower and higher values of $U_c$. In order to clarify this effect, we plot the ASR versus $\lambda_d$ in a 2-D plot with $U_c \in \{1,14\}$ equivalent to $T_c \in \{5,70\}$ in \figref{fig:ASR_ld_uc-1-14_phi5}.

As seen in \figref{fig:ASR_ld_uc-1-14_phi5}, for $U_c = 1$ user and $T_c =5$ antennas, the rate contributed from the CUEs to the sum rate is low as there is only one CUE. This rate is in a comparable level as the contribution of D2D users sum rate to the total ASR. Adding D2D users to the network (i.e., increasing $\lambda_d$), which may cause interference, will nevertheless leads to an increase in the ASR. This increase in the ASR continues until reaching a certain density that gives the maximum ASR. %
By further increasing $\lambda_d$, the interference between D2D users reduces their coverage probability  as previously observed in Remark~\ref{remark:P_succ_d2d_lambda_d}. This limits the per link data rate and even a high number of D2D users cannot compensate for the D2D rate loss. %
At the same time, increasing $\lambda_d$ tremendously affects the CUEs sum rate (cf.\ Remark~\ref{remark:P_succ_cue_lambda_dinf}). Consequently, as $\lambda_d$ increases, the ASR decreases.

By increasing the number of CUEs and BS antennas to $U_c = 14$ users and $T_c = 70$ antennas, respectively, in \figref{fig:ASR_ld_uc-1-14_phi5}, the average rates of the CUEs become higher than the case with $U_c = 1$ user and $T_c =5$ antennas as expected from Corollary~\ref{corollary:P_succ_cue_TcUcinf} and the multiplexing gain from having many CUEs. However, by introducing a small number of D2D users, there is a substantial probability that the interference from the D2D users reduces the CUEs' rates per link as observed in Remark~\ref{remark:P_succ_cue_lambda_dinf}. The reduction in these rates are not compensated in the ASR by the contribution of the D2D users' rates. Note that, as we stated in Remark~\ref{remark:P_succ_d2d_Tc_Uc}, when $U_c$ is scaled with $T_c$, it impacts the D2D coverage probability, but the decrease in the performance of D2D users is not significant. Furthermore, if we keep increasing $\lambda_d$, even though the rate per link decreases for both CUEs and D2D users, there is a local minima after which the aggregate D2D rate over all D2D users  becomes higher and the ASR increases again. The second turning point follows from the same reasoning as for the case of $U_c=1$ user and $T_c = 5$ antennas, i.e., in higher D2D densities, the interference from D2D users are the limiting factor for the ASR. This effect can also be observed in \figref{fig:ASR_ld-4-6_uc_phi5} where the ASR performance is depicted versus different number of CUEs (and BS antennas) for two D2D densities. At the lower density, the ASR is linearly increasing with $U_c$ (and $T_c)$, however, in the interference-limited regime (higher $\lambda_d$), increasing the number of CUEs and BS antennas do not impact the network ASR performance.

\begin{figure}[tp]
\centering
\includegraphics[width=0.65\columnwidth]{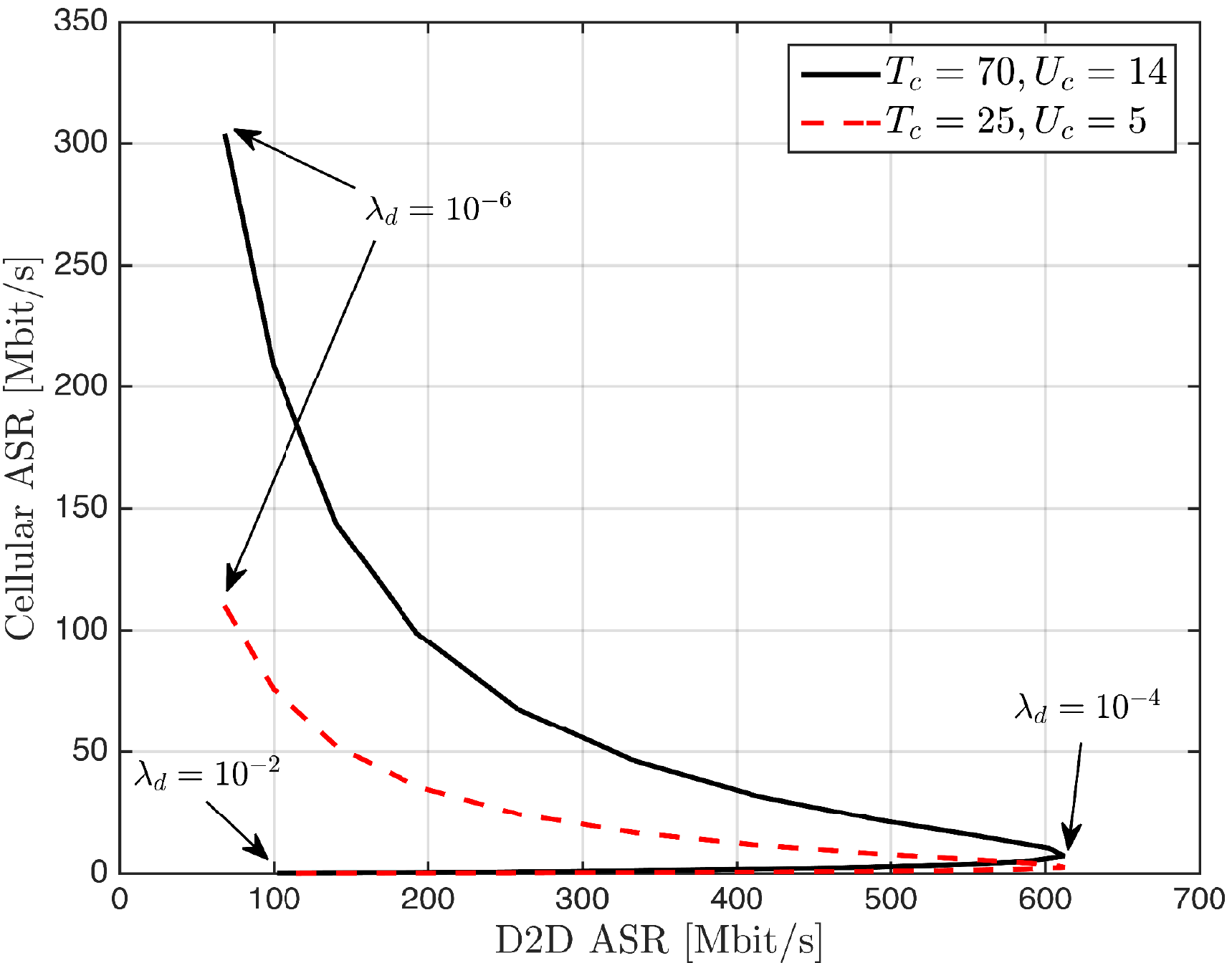}
\caption{Cellular ASR vs. D2D ASR $\mathrm{[Mbit/s]}$ for a fixed ratio $\frac{T_c}{U_c}= 5$. The curves are obtained by varying the value of $\lambda_d$ from $10^{-6}$ to $10^{-2}$.}
\label{fig:asrC_asrD_ld}
\end{figure}

The reasoning in \figref{fig:ASR_ld_uc-1-14_phi5} and \figref{fig:ASR_ld-4-6_uc_phi5} can be well understood from \figref{fig:asrC_asrD_ld} which explains the tradeoff between the ASR of CUEs and D2D users in the network. In the scenario  in which we have $T_c = 70$ antennas and $U_c =14$ users, the cellular network contributes more to the total ASR for the low D2D density regime (e.g., $\lambda_d=10^{-6}$) due to high number of CUEs and BS antennas. In this region, the ASR gains from massive MIMO is large. By increasing $\lambda_d$, the gain from massive MIMO vanishes as the interference added by the D2D users  dominates and degrades the performance that was achieved by interference cancellation between CUEs. %
Therefore, with medium D2D user density, if there is a fixed rate constraint for CUEs, the network can still benefit (from the ASR perspective) from underlay D2D communications. However, in the high D2D density regime (e.g., $\lambda_d = 10^{-4}$), the cellular ASR is too small and it is better that the cellular and D2D tiers use the overlay approach for communication instead of the underlay approach.

\begin{figure}[tp]
\centering
\includegraphics[width=0.65\columnwidth]{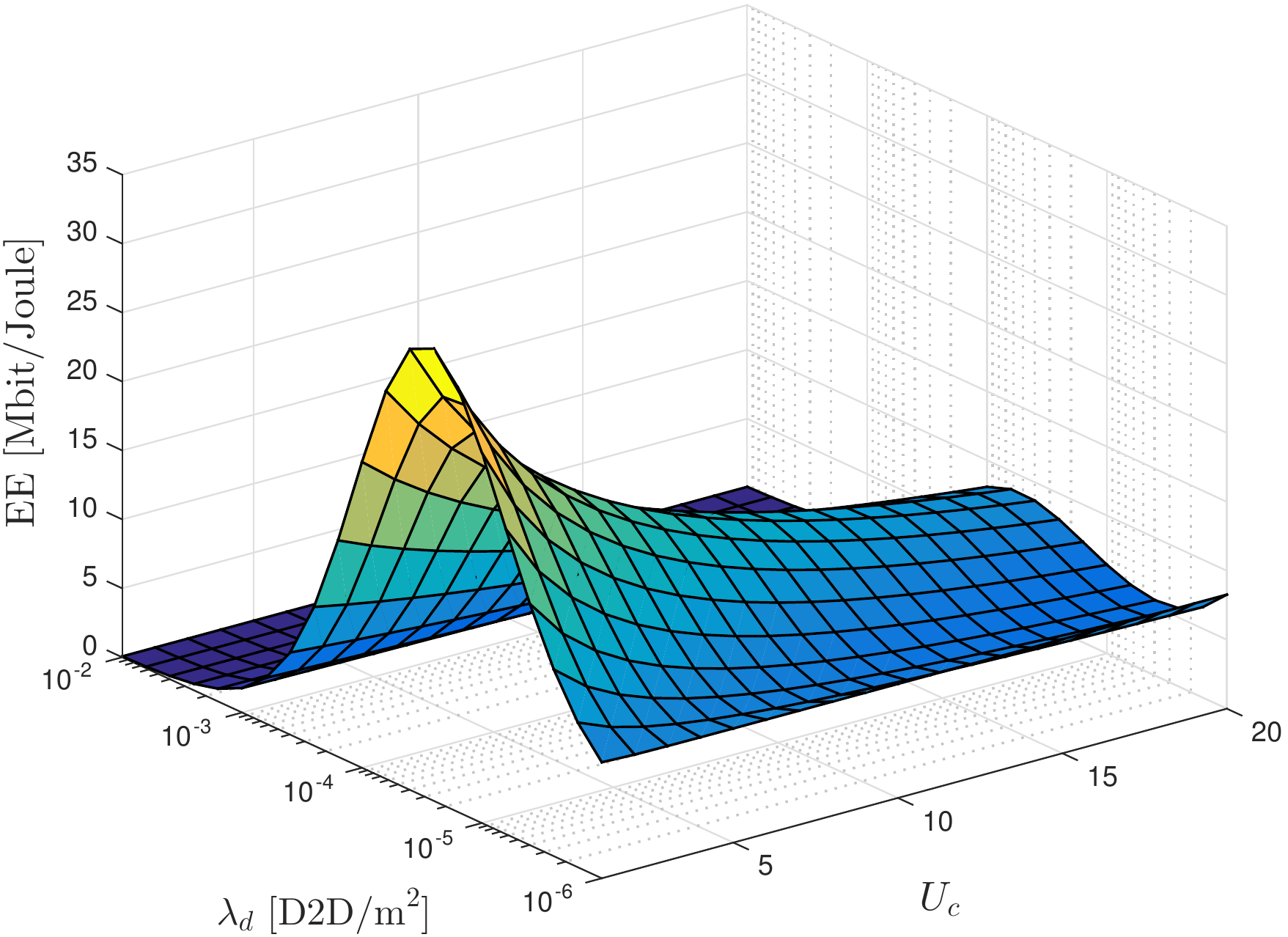}
\caption{EE $\mathrm{[Mbit/Joule]}$ as a function of the number of CUEs $U_c$ and the D2D user density $\lambda_d$ for a fixed ratio $\frac{T_c}{U_c}= 5$.}
\label{fig:EE_ld_uc_phi5}
\end{figure}

\begin{figure*}[tp]
\centering
\subfloat[]{
\includegraphics[width=.48\linewidth]{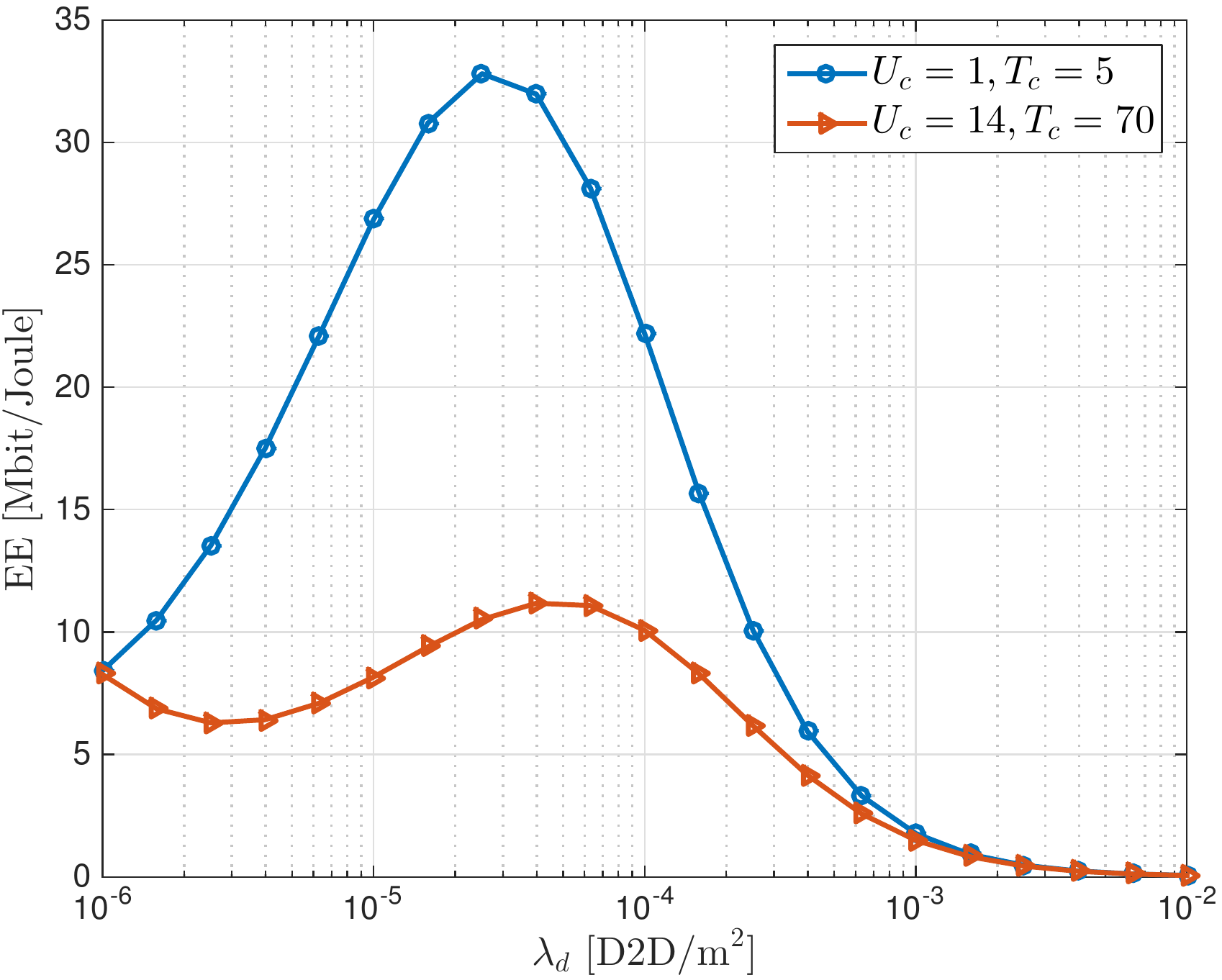}
\label{fig:EE_ld_uc-1-14_phi5}
}
\hfil
\subfloat[]{
\includegraphics[width=.48\linewidth]{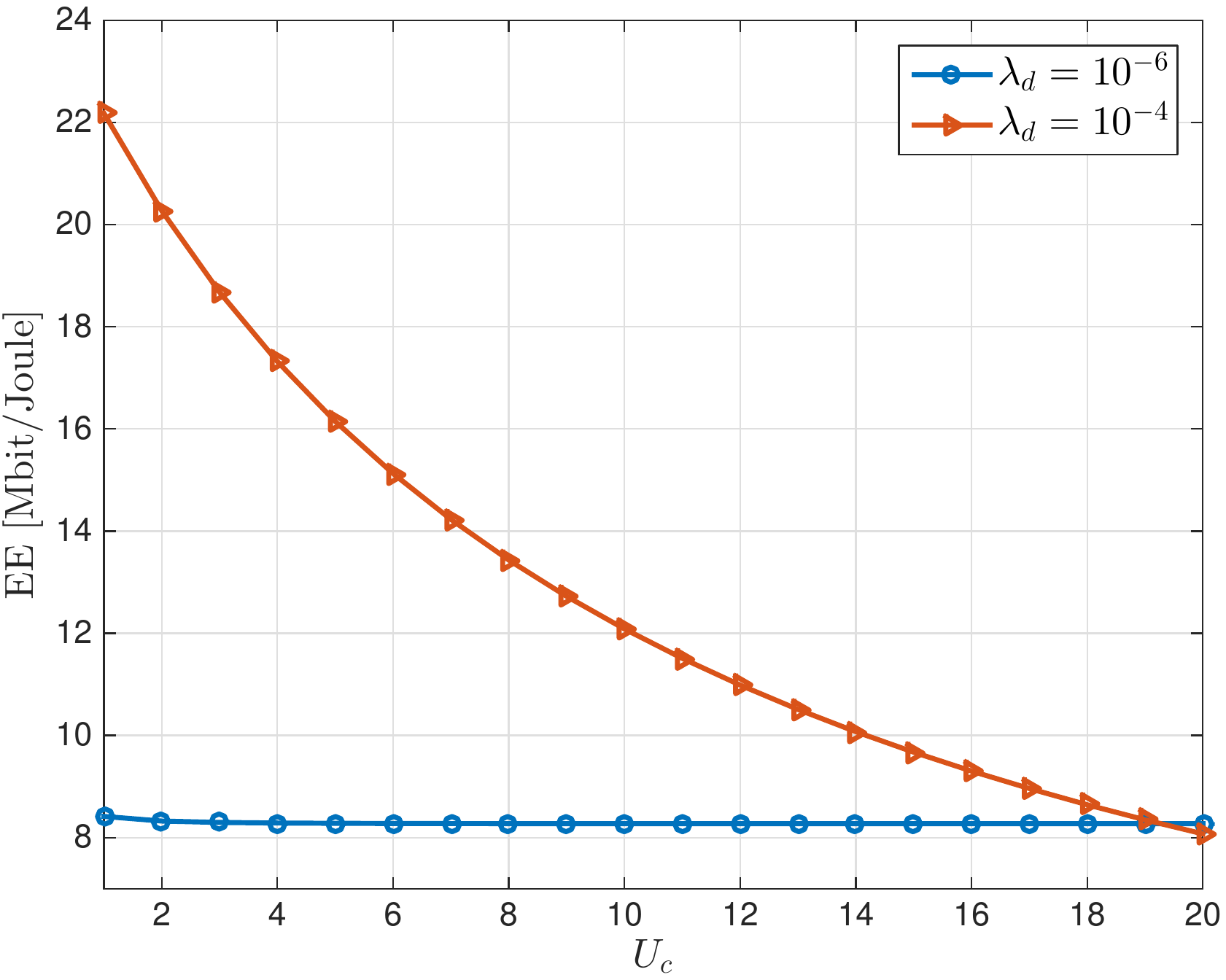}
\label{fig:EE_ld-4-6_uc_phi5}
}
\caption{EE $\mathrm{[Mbit/Joule]}$: (a) as a function of the D2D user density $\lambda_d$ for a fixed ratio $\frac{T_c}{U_c}= 5$ with the number of CUEs $U_c\in\{1,14\}$; (b) as a function of the number of CUEs $U_c$ with the D2D user density $\lambda_d\in\{10^{-6},10^{-4}\}$ for a fixed ratio  $\frac{T_c}{U_c}= 5$.}
\label{fig:EE_2D}
\end{figure*}

In \figref{fig:EE_ld_uc_phi5}, we show the network performance in terms of the EE as a function of the parameters $\lambda_d$ and $U_c$ with $\frac{T_c}{U_c}= 5$. It is observed that the EE is a decreasing function of $U_c$ and $T_c$. In contrast, there is a maximum point in the EE based on different values of $\lambda_d$. To study this result further, similar to the ASR, we first plot the EE versus $\lambda_d$ for $U_c \in \{1,14\}$ and $T_c \in \{4,70\}$ in \figref{fig:EE_ld_uc-1-14_phi5}. We can see that the pattern for both low and high number of BS antennas are similar to \figref{fig:ASR_ld_uc-1-14_phi5}. The higher EE is achieved with $U_c =1$ user and $T_c=5$ antennas as opposed to $U_c =14$ users and $T_c=70$ antennas. This is because the extra circuit power of the cellular tier with $U_c =14$ users and $T_c=70$ antennas does not bring any substantial ASR improvement over the case with  $U_c =1$ user and $T_c=5$ antennas. %

Furthermore, if we plot the EE versus $U_c$, we see a different behavior for low and high D2D densities. \figref{fig:EE_ld-4-6_uc_phi5} illustrates that in the low D2D density regime ($\lambda_d=10^{-6}$), even though the ASR increases linearly, the EE almost stays the same as the number of CUEs, and correspondingly the number of BS antennas, increases. From \eqref{eq:P_tot}, we can observe that for a fixed $\lambda_d$, only the circuit power is changed by increasing $U_c$ and $T_c$. At the same time, the circuit power dominates the the total power consumption and increases almost linearly leading to an (almost) constant EE. %
The network performance in terms of the EE is poor with high density of D2D users ($\lambda_d=10^{-4}$). This is due to the fact that the sum rate contributed by the CUEs is already degraded by the interference from high number of D2D users, and additionally, increasing $U_c$ (and accordingly $T_c$) increases the circuit power without any gain in the total ASR. Consequently, the EE decreases. Thus, massive MIMO can only improve the EE if the D2D user density is small, otherwise dedicated resources or underlaying with fewer BS antennas is beneficial.

\subsection{Fixed Number of CUEs}

\begin{figure*}[tp]
\centering
\subfloat[]{
\includegraphics[width=.48\linewidth]{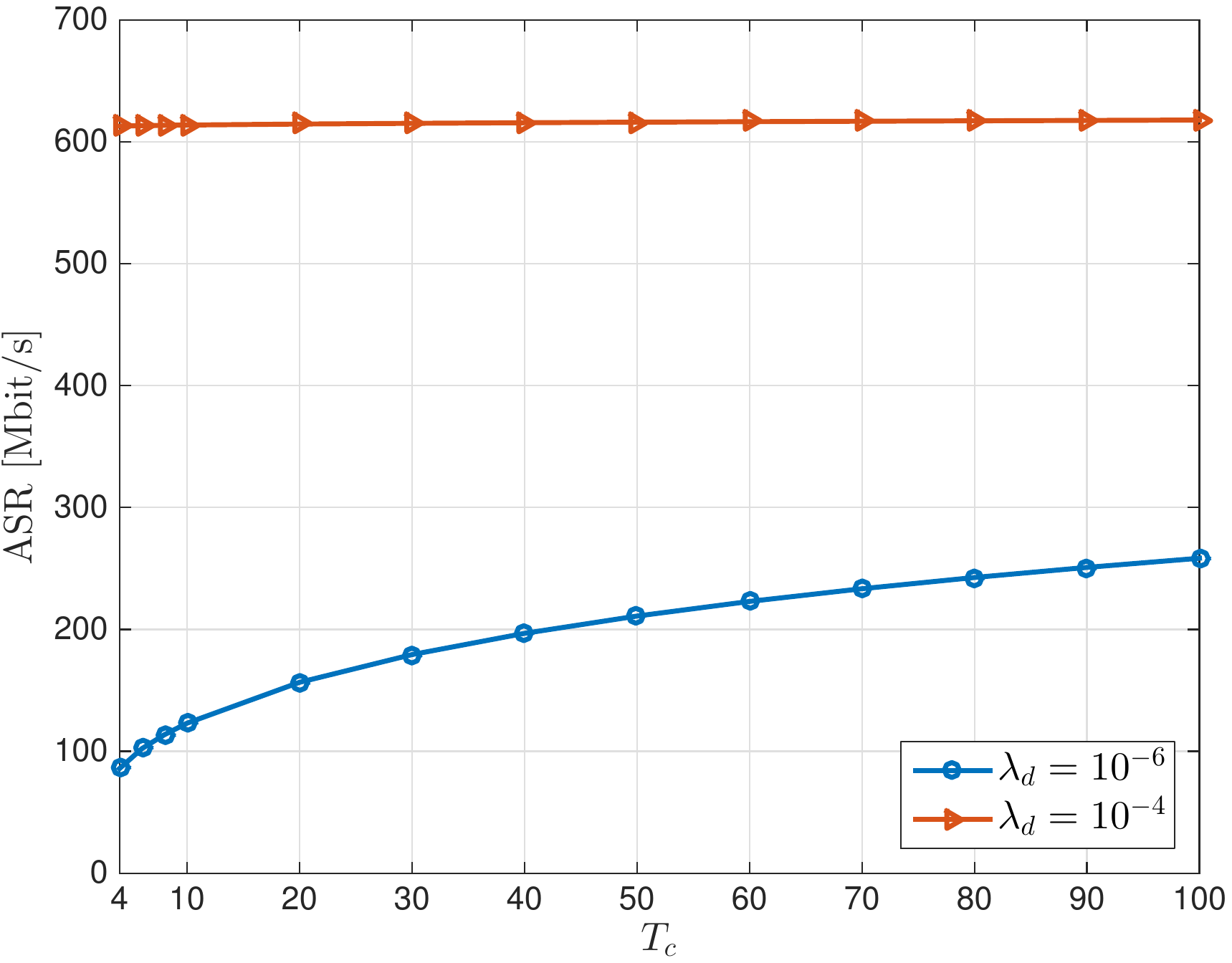}
\label{fig:ASR_ld-4-6_tc}
}
\hfil
\subfloat[]{
\includegraphics[width=.48\linewidth]{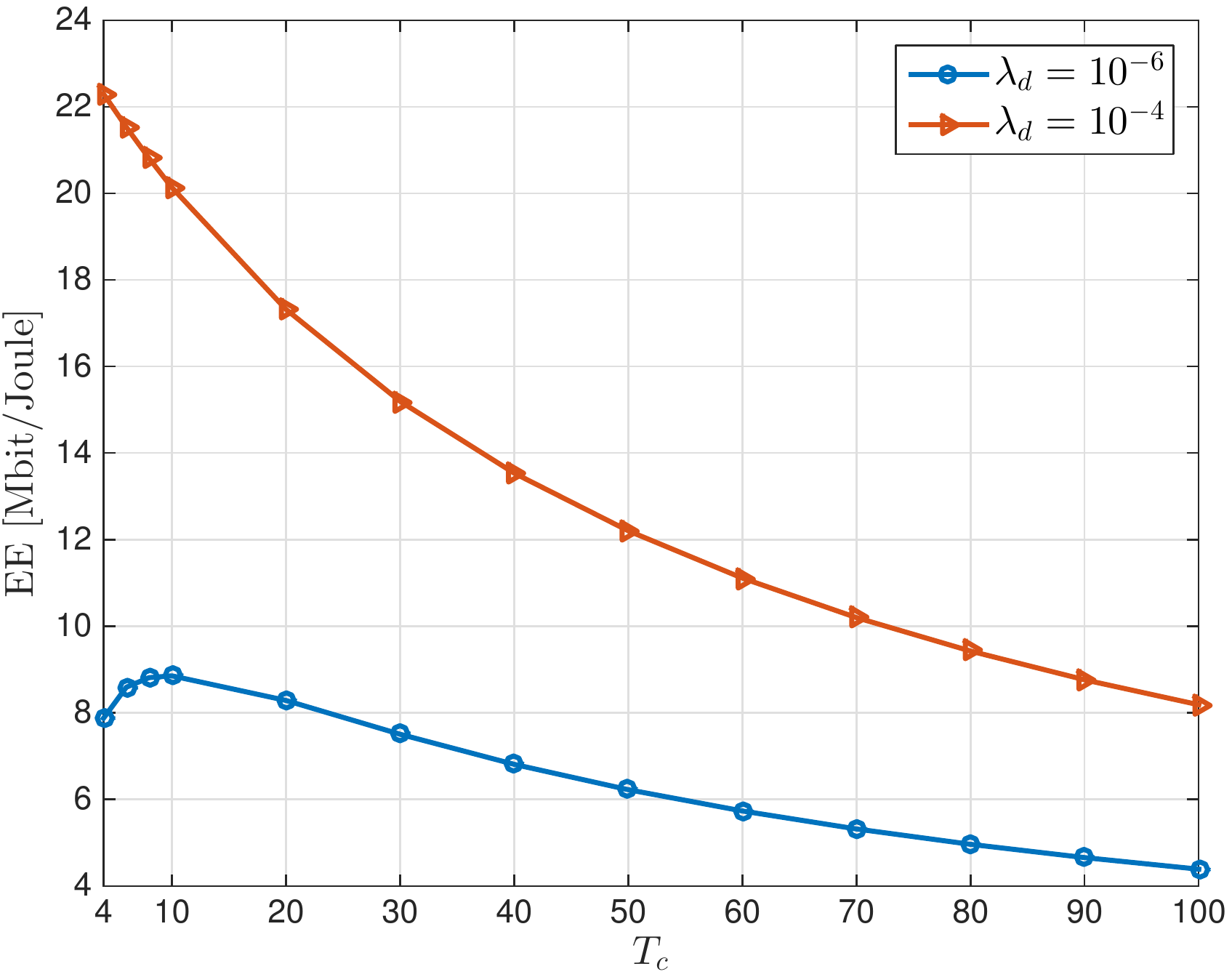}
\label{fig:EE_ld-4-6_tc}
}
\caption{(a) ASR $\mathrm{[Mbit/s]}$ and (b) EE $\mathrm{[Mbit/Joule]}$ as a function of the number of BS antennas $T_c$ for $U_c=4$ users and  $\lambda_d \in \{10^{-6},10^{-4}\}$.}
\label{fig:Uc4_2D}
\end{figure*}

In this section, we evaluate the system performance when the number of CUEs is fixed with $U_c = 4$ users. The general trend of the network performance is the same as the case with $\frac{T_c}{U_c} =5$ in the previous section. However, there are some differences which are highlighted in \figref{fig:ASR_ld-4-6_tc} and \figref{fig:EE_ld-4-6_tc} for the ASR and EE, respectively. As it is shown in \figref{fig:ASR_ld-4-6_tc}, in the low D2D user density regime (i.e., $\lambda_d=10^{-6}$) the ASR is increasing in $T_c$, however, with a lower slope as compared to the case of $\frac{T_c}{U_c} =5$. By increasing the number of BS antennas for the fixed number of CUEs, better performance per user can be achieved, however in this case,  as the number of CUEs is not high, the ASR increases with a small slope. %
For high D2D user density (i.e., $\lambda_d = 10^{-4}$), the ASR is almost flat.

\figref{fig:EE_ld-4-6_tc} illustrates that when the D2D user density is low, the EE benefits from adding extra BS antennas until the sum of the circuit power consumption of all antennas dominates the performance and leads to a gradual decrease in the EE. As the figure implies, there exists an optimal number of BS antennas which is relatively small since the main massive MIMO gains come from multiplexing rather than just having many antennas. However, in high density D2D scenario, which is the interference-limited scenario, the EE decreases monotonically with $T_c$. Increasing the number of BS antennas in this region cannot improve the ASR significantly, as shown in \figref{fig:ASR_ld-4-6_tc}; at the same time the circuit power consumption increases as a result of the higher number of BS antennas, which in turn leads to decreasing network EE.

The conclusion is that the D2D user density has a very high impact on a network that employs the massive MIMO technology. In the downlink, these two technologies can only coexist in low density of D2D users with careful interference coordination. The number of CUEs should be a function of the number of BS antennas in order to benefit from high number of BS antennas in terms of the ASR and EE. Otherwise, in high density of D2D users, the D2D communication should use the overlay approach rather than the underlay, that is, dedicated time/frequency resources should be allocated to the D2D tier.

\subsection{The Effect of Other System Parameters}

So far, we have discussed the results based on constant transmit power $P_c$,  D2D transmit power $P_d$, and distance between D2D~Tx-Rx pairs $R_{0,0}$ given in Table~\ref{table:Sim_param}. Now we comment on the choice of these parameters and study their effects on the system performance. From Proposition~\ref{proposition:P_succ_d2d}, Proposition~\ref{proposition:P_succ_cue}, and Remark~\ref{remark:P_succ_d2d_PcPd}, it is evident that the coverage probability for both D2D and cellular tiers, and consequently the network ASR and EE, depend on the ratio of $P_d$ and $P_c$. Therefore, we fix $P_c$ and vary $P_d$.

\begin{figure*}[tp]
\centering
\subfloat[]{
\includegraphics[width=.48\linewidth]{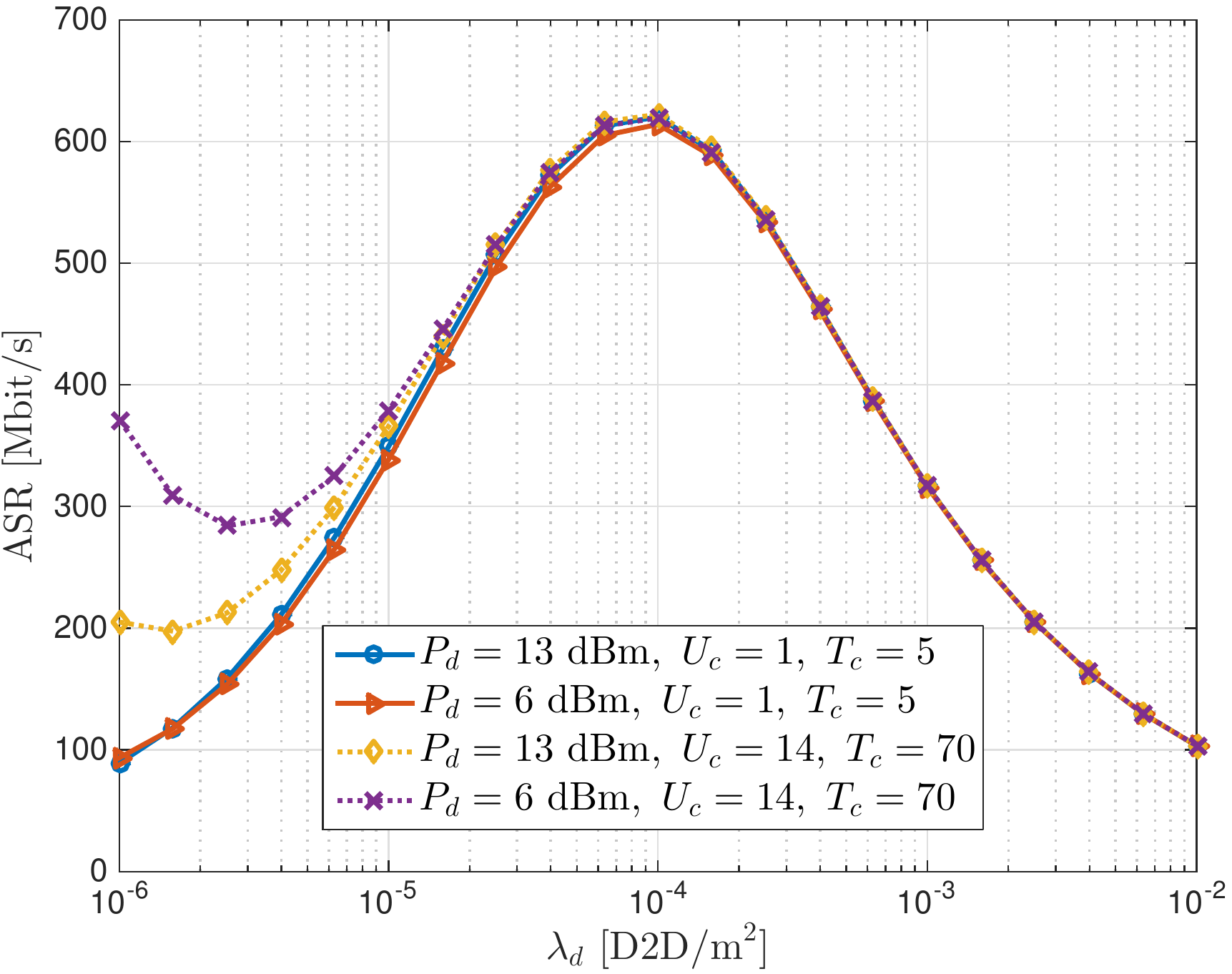}
\label{fig:ASR_ld_phi5-diff-Pd}
}
\hfil
\subfloat[]{
\includegraphics[width=.48\linewidth]{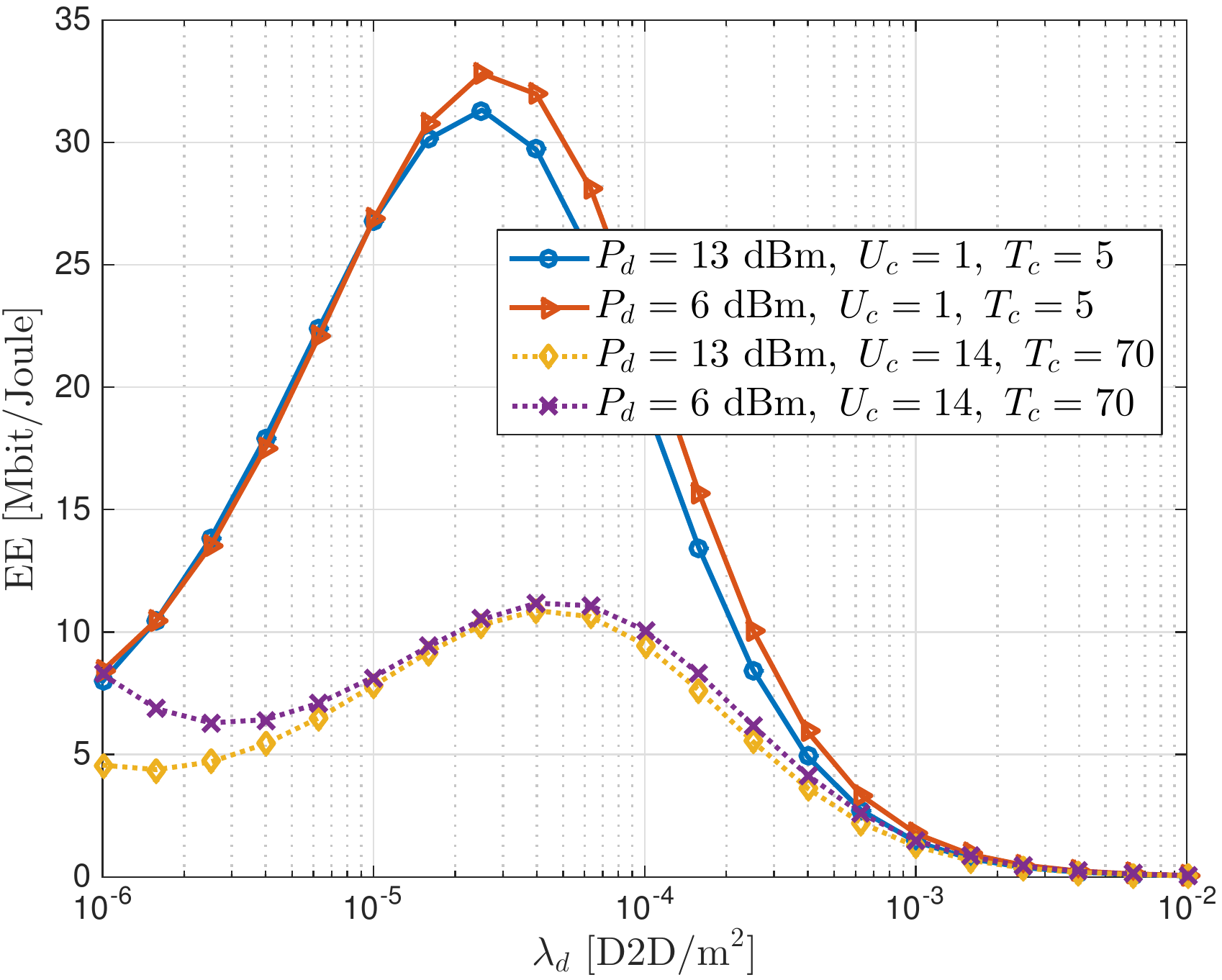}
\label{fig:EE_ld_phi5-diff-Pd}
}
\caption{(a) ASR $\mathrm{[Mbit/s]}$ and (b) EE $\mathrm{[Mbit/Joule]}$ as a function of the D2D user density $\lambda_d$ for different D2D transmit power and a fixed ratio $\frac{T_c}{U_c}= 5$ with the number of CUEs $U_c\in\{1,14\}$.}
\label{fig:Phi5_Pd}
\end{figure*}

\figref{fig:ASR_ld_phi5-diff-Pd} shows the ASR as function of $\lambda_d$ under two different power levels, i.e., $P_d = 6~\mathrm{dBm}$ and $P_d = 13~\mathrm{dBm}$ in a scenario where the number of CUEs $U_c$ is scaled by $T_c$. We see that higher $P_d$ degrades the ASR at higher number of CUEs (and BS antennas) when the D2D user density is low, but has negligible impact at lower number of CUEs. The reason is that  increasing $P_d$, on the one hand, boosts the D2D user rates, and on the other hand, causes more interference to CUEs which deteriorates their rates. Consequently, at low D2D user densities and high number of CUEs and BS antennas where the cellular sum rate is the main contributer to the total ASR, the interference caused by higher D2D transmit power is the dominant factor leading to lower total ASR. However, as $\lambda_d$ increases, the contribution of the D2D sum rate to the total ASR increases, and thus with higher $P_d$, the increase in the D2D sum rates compensates the decrease in CUEs sum rate and the difference in terms of the total ASR between the different power levels vanishes. When the number of CUEs is small, i.e., $U_c=1$ user and $T_c=5$ antennas, the CUE and D2D users have almost the same contributions to the ASR and increasing $P_d$ has negligible impact on the performance.

\figref{fig:EE_ld_phi5-diff-Pd} depicts the EE as a function of $\lambda_d$ under the same two levels of D2D transmit power. It is observed that lower $P_d$ is more beneficial in terms of the EE in both cases of $U_c = 1$ user and $U_c=14$ users. This is particularly visible in higher density of D2D users (e.g., $\lambda_d=3\times 10^{-5}$) with $U_c = 1$ user and $T_c=5$ antennas when the interference is the limiting factor. %
With $U_c = 14$ users and $T_c=70$ antennas, the CUEs have higher impact on the ASR, and as a consequence, the system benefits from lower transmit power of D2D users in terms of the EE. Therefore, we have chosen $P_d = 6~\mathrm{dBm}$ in the previous performance evaluation, as it has a better impact on the ASR as well as EE, especially in higher number of BS antennas.

\begin{figure}[tp]
\centering
\includegraphics[width=0.65\columnwidth]{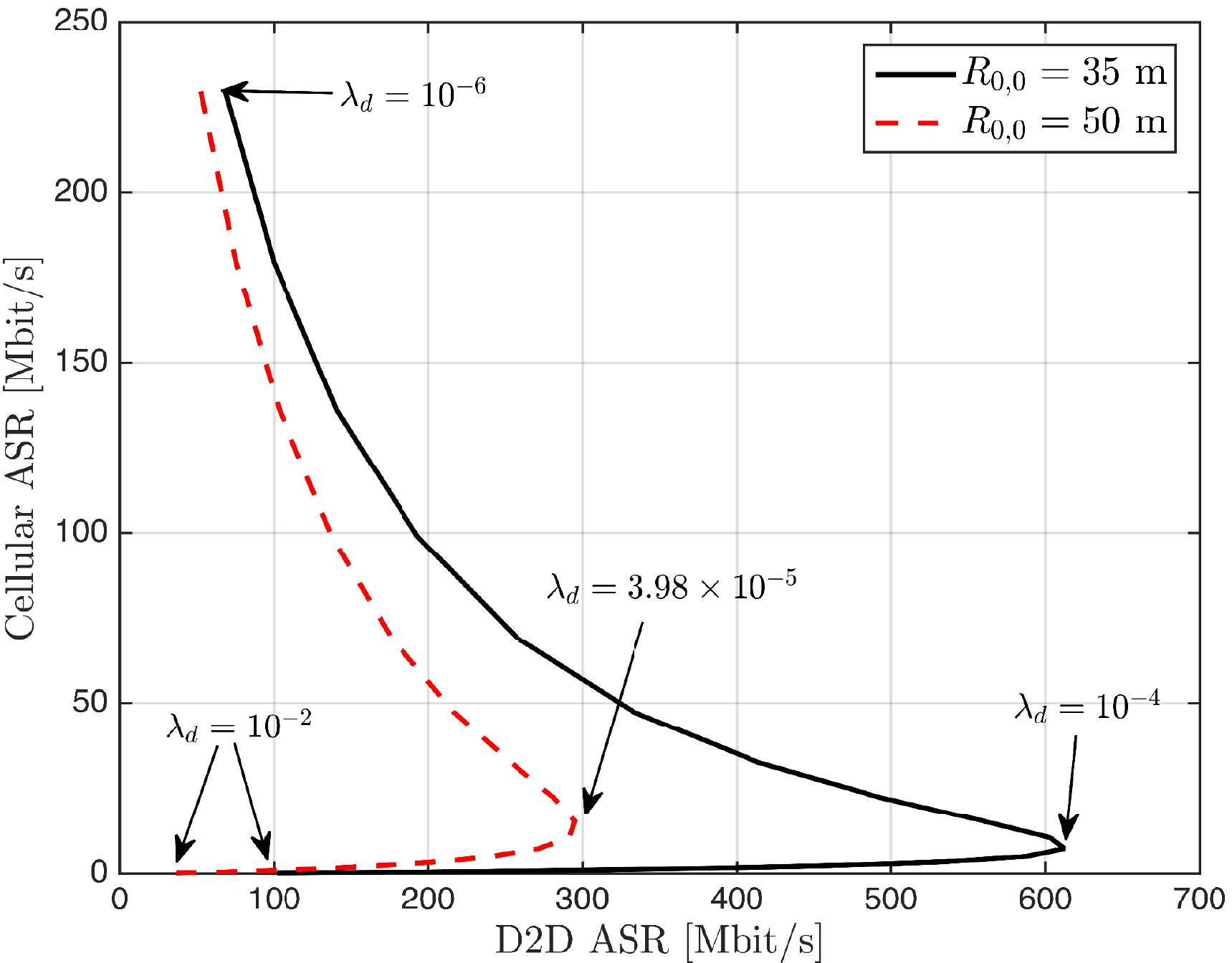}
\caption{Cellular ASR vs. D2D ASR $\mathrm{[Mbit/s]}$ for different distances between D2D~Tx and D2D~Rx with $U_c=4$ users and $T_c=70$ antennas. The curves are obtained by varying the value of $\lambda_d$ from $10^{-6}$ to $10^{-2}$.}
\label{fig:asrC_asrD_diff-d2d-distance}
\end{figure}

Another important parameter that impacts the ASR is the distance between D2D~Tx-Rx pairs, i.e., $R_{0,0}$. The effect of this parameter is only on the coverage probability of D2D users as seen in Proposition~\ref{proposition:P_succ_d2d} and Proposition~\ref{proposition:P_succ_cue}. \figref{fig:asrC_asrD_diff-d2d-distance} illustrates the cellular ASR versus the D2D ASR for different values of $\lambda_d$ and $R_{0,0}$. The figure verifies that by decreasing $R_{0,0}$ only the ASR of D2D tier increases and as Remark~\ref{remark:P_succ_d2d_Uc} implies increasing $R_{0,0}$ decreases the coverage probability of D2D users leading to lower ASR and EE. Since D2D communications are mostly meant for close proximity applications, we have chosen $R_{0,0} =35~\mathrm{m}$ in our performance study. Moreover, by decreasing the distance between D2D users, more D2D users can coexist simultaneously. This is observed in \figref{fig:asrC_asrD_diff-d2d-distance} that with $R_{0,0} =35~\mathrm{m}$ the maximum ASR (of the D2D tier as well as the network) is achieved at the D2D density $\lambda_d=10^{-4}$ while with $R_{0,0} =50~\mathrm{m}$, it is achieved at the D2D density $\lambda_d=3.98 \times 10^{-5}$.

\section{Conclusions}
\label{sec:conclusion}
We studied the coexistence of two key 5G concepts: device-to-device (D2D) communication and massive MIMO. We considered two performance metrics, namely, the average sum rate in $\mathrm{bit/s}$ and the energy efficiency in $\mathrm{bit/Joule}$. %
We considered a setup with a number of uniformly distributed cellular users in the cell, while the D2D transmitters are distributed according to a Poisson point process. We derived tractable expressions for the coverage probabilities of both cellular and D2D users which led to computation of the average sum rate and energy efficiency. We then studied the tradeoff between the number of base station antennas, the number of cellular users, and the density of D2D users for a given coverage area in the downlink. Our results showed that both the average sum rate and energy efficiency behave differently in scenarios with low and high density of D2D users. %

Underlay D2D communications and massive MIMO can only coexist in low densities of D2D users with careful interference coordination, because the massive MIMO gains vanish when the interference from the D2D tier becomes too large. The number of cellular users should be a function of the number of base station antennas in order to benefit from high number of base station antennas in terms of the average sum rate and energy efficiency. If there is a high density of D2D users, the D2D communication should use the overlay approach rather than the underlay or the network should only allow a subset of the D2D transmissions to be active at a time.

\appendices

\section{Proof of Proposition~\ref{proposition:P_succ_d2d}}
\label{sec:proof_Pcov_d2d}

The proof follows by substituting the definition of $\mathrm{SINR}_d$ from \eqref{eq:SINR_d2d} into \eqref{eq:PsuccDef} where we obtain
\begin{align}
\mathrm{P}^d_{\mathrm{cov}}(\beta_d) &= \mathrm{Pr}\big\{\mathrm{SINR}_d \geq \beta_d \big\} \nonumber \\
&= \mathrm{Pr} \left\{ P_{d}  R_{0,0}^{-\alpha_d} |g_{0,0}|^2 \geq \beta_d \Big(I_{\textrm{BS},0} + I_{d,0} +  \frac{N_0}{A_d }\Big) \right\} \nonumber \\
&= \mathrm{Pr} \left\{ |g_{0,0}|^2 \geq  \frac{\beta_d}{P_dR_{0,0}^{-\alpha_d}}   \Big(I_{\textrm{BS},0} + I_{d,0} +  \frac{N_0}{A_d }\Big) \right\} \nonumber \\
&\eqtop{a} \mathbb{E}_{I_{\textrm{BS},0}, I_{d,0}} \left[\exp\bigg(- \frac{\beta_d}{P_dR_{0,0}^{-\alpha_d}} \Big(I_{\textrm{BS},0} + I_{d,0} +  \frac{N_0}{A_d }\Big)\bigg)\right]\nonumber \\
&\eqtop{b}   \mathbb{E}_{I_{\textrm{BS},0}}\left[\exp\bigg(- \frac{\beta_d}{P_dR_{0,0}^{-\alpha_d}} I_{\textrm{BS},0}\bigg)\right] \mathbb{E}_{I_{d,0}}\left[\exp\bigg(- \frac{\beta_d}{P_dR_{0,0}^{-\alpha_d}} I_{d,0}\bigg)\right]  \exp\bigg(-\frac{\beta_d }{\bar{\gamma}_d}\bigg) \nonumber \\
&\eqtop{c} \mathcal{L}_{I_{\textrm{BS},0}}\bigg(\frac{\beta_d}{P_dR_{0,0}^{-\alpha_d}}\bigg)\mathcal{L}_{I_{d,0}}\bigg(\frac{\beta_d}{P_dR_{0,0}^{-\alpha_d}}\bigg) \exp\bigg(-\frac{\beta_d}{\bar{\gamma}_d}\bigg). \label{eq:Pd2dProof}
\end{align}
Step $(a)$ comes from the fact that $|g_{0,0}|^2 \sim \exp(1)$ and $(b)$ follows since the noise and interference terms are mutually independent. In step $(c)$, the Laplace transform defined as $\mathcal{L}_{x}(s) = \mathbb{E}_{x}\big[e^{- sx}\big]$ is identified.

The first Laplace transform in \eqref{eq:Pd2dProof} is with respect to $I_{\textrm{BS},0}$ in \eqref{eq:I_BS0} which is a function of two random variables, namely $\|\vect{f}_{0,\textrm{BS}}^H \mathbf{V}\|^2$ and $R_{0,\textrm{BS}}$. This Laplace transform is calculated as
\begin{align}
\mathcal{L}_{I_{\textrm{BS},0}}&\bigg(\frac{\beta_d}{P_dR_{0,0}^{-\alpha_d}}\bigg) = \mathbb{E}_{I_{\textrm{BS},0}} \left[\exp\bigg( - \frac{\beta_d}{P_dR_{0,0}^{-\alpha_d}} I_{\textrm{BS},0}\bigg)\right]\nonumber \\
&= \mathbb{E}_{R_{0,\textrm{BS}}} \left[\mathbb{E}_{\|\vect{f}_{0,\textrm{BS}}^H \mathbf{V}\|^2}\bigg[\exp\bigg(- \frac{\beta_d}{P_dR_{0,0}^{-\alpha_d}} \frac{\zeta R_{0,\textrm{BS}}^{-\alpha_c}}{A_d}\|\vect{f}_{0,\textrm{BS}}^H \mathbf{V}\|^2 \bigg) \Big| R_{0,\textrm{BS}}\bigg] \right] \nonumber \\
&=\mathbb{E}_{R_{0,\textrm{BS}}} \left[  \mathcal{L}_{\|\vect{f}_{0,\textrm{BS}}^H \mathbf{V}\|^2}\bigg(\frac{\beta_d}{P_dR_{0,0}^{-\alpha_d}} \frac{\zeta R_{0,\textrm{BS}}^{-\alpha_c}}{A_d}\bigg)\right]\nonumber \\
&\eqtop{a} \mathbb{E}_{R_{0,\textrm{BS}}} \left[\frac{1}{(\kappa \beta_d R_{0,\textrm{BS}}^{-\alpha_c} + 1)^{U_c}}\right]\nonumber \\
&\eqtop{b} \int_0^R \frac{2r}{R^2(\kappa \beta_d r^{-\alpha_c}+1)^{U_c}} \mathrm{d}r\nonumber \\
&\eqtop{c}  \frac{2(\kappa\beta_d)^{2/\alpha_c}}{\alpha_c R^2}\int_0^{y} \frac{t^{U_c+\frac{2}{\alpha_c}-1}}{(1-t)^{\frac{2}{\alpha_c}+1}}\mathrm{d}t \nonumber \\
&\eqtop{d} \frac{ (\kappa\beta_d)^{2/\alpha_c}}{R^2}\left(y^{U_c + \frac{2}{\alpha_c}-1} (1-y)^{- \frac{2}{\alpha_c}}- \Big(U_c + \frac{2}{\alpha_c}-1\Big) \mathcal{B}\Big(y; U_c + \frac{2}{\alpha_c}-1, 1-\frac{2}{\alpha_c}\Big) \right) \label{eq:LT_I_BS0}
\end{align}
for $\alpha_c>2$, where $(a)$ follows by introducing the notation
\begin{equation}
\kappa =  \frac{\zeta}{P_d A_d R_{0,0}^{-\alpha_d}}
\end{equation}
and from the Laplace transform of the  probability density function (PDF) of $\|\vect{f}_{0,\textrm{BS}}^H \mathbf{V}\|^2$ which, by neglecting the spatial correlation, is tightly approximated by a Chi-squared distribution  as $2\|\vect{f}_{0,\textrm{BS}}^H \mathbf{V}\|^2 \sim \chi^2_{2U_c}$ \cite{Dhillon-2013-TWC}. Note that
\begin{align}\label{eq:normfHv}
\big\|\vect{f}_{0,\textrm{BS}}^H \mathbf{V}\big\|^2 &= \big\|\vect{f}_{0,\textrm{BS}}^H [\vect{v}_{0},\ldots,\vect{v}_{{U_c}-1}]\big\|^2 \nonumber\\
&=\sum_{i=0}^{U_c-1} |\vect{f}_{0,\textrm{BS}}^H\vect{v}_{i}|^2,
\end{align}
where $\vect{f}_{0,\textrm{BS}}^H\vect{v}_{i}$, $i = \{0, \ldots, U_c-1\}$, are zero-mean circular symmetric complex Gaussian random variables with unit variance. Therefore,  $\sum_{i=0}^{U_c-1} |\vect{f}_{0,\textrm{BS}}^H\vect{v}_{i}|^2$ is the summation of $U_c$ i.i.d.\ exponential random variables which has an $\mathrm{Erlang}(U_c,1)$ distribution. Equivalently, the sum scaled down by $\frac{\sigma^2}{2}$ (i.e., multiplied by $\frac{2}{\sigma^2}$) has a (standard) Chi-squared distribution with $2U_c$ degrees of freedom. Hence, the PDF of $\|\vect{f}_{0,\textrm{BS}}^H \mathbf{V}\|^2$ is
\begin{equation}\label{eq:pdf_IBS0}
f_{\|\vect{f}_{0,\textrm{BS}}^H \mathbf{V}\|^2} (x) =  \frac{x^{Uc-1}e^{-x}}{(U_c-1)!}.
\end{equation}
From Laplace transform theory we know that  $\mathcal{L}\big[t^n e^{-\alpha t}\big] = \frac{n!}{(s+\alpha)^{n+1}}$ and with some simplifications, we obtain the result in step $(a)$. Step $(b)$ in \eqref{eq:LT_I_BS0} follows from the PDF of $R_{0,\textrm{BS}}$ which is
\begin{equation}
f_{R_{0,\textrm{BS}}}(r) = \left\{\begin{array}{ll}
\frac{2r}{R^2}, & \textrm{if } 0\leq r \leq R,\\
0, & \textrm{otherwise}, \end{array} \right.
\end{equation}
as the typical D2D~Rx is uniformly distributed over the cell area and the BS is located in the cell center. %
Step $(c)$ in \eqref{eq:LT_I_BS0} is obtained by the change of variable $\frac{1}{\kappa \beta_d r^{-\alpha_c} + 1}\rightarrow t$ which leads to the integral boundary $y \triangleq \frac{1}{\kappa \beta_d R^{-\alpha_c} + 1} $. Finally, $(d)$ follows by integration by part where $\mathcal{B}(x;a,b)$ is the incomplete Beta function defined as
\begin{equation}\label{eq:beta_func}
\mathcal{B}(x;a,b) = \int_0^x t^{a-1} (1-t)^{b-1} \mathrm{d}t,
\end{equation}
for $a,b>0$.

Next, we proceed to calculate the second Laplace transform in \eqref{eq:Pd2dProof}. This transform is with respect to $I_{d,0}$ in \eqref{eq:I_d0} which is a function of two random variables, that is $|g_{0,j}|^2$ and $R_{0,j}$. Therefore, we have
\begin{align}
\mathcal{L}_{I_{d,0}} \bigg(\frac{\beta_d}{P_dR_{0,0}^{-\alpha_d}}\bigg) &= \mathbb{E}_{I_{d,0}}\left[\exp\Big(- \frac{\beta_d}{P_dR_{0,0}^{-\alpha_d}}  I_{d,0}\Big)\right]\nonumber \\
&= \mathbb{E}_{R_{0,j},|g_{0,j}|^2} \left[\exp\Big(- \frac{\beta_d}{P_dR_{0,0}^{-\alpha_d}} \sum_{j\neq 0} P_d R_{0,j}^{-\alpha_d} |g_{0,j}|^2\Big)\right] \nonumber \\
&=  \mathbb{E}_{R_{0,j}} \left[\prod_j  \mathbb{E}_{|g_{0,j}|^2}\bigg[\exp\Big(-\frac{\beta_d}{R_{0,0}^{-\alpha_d}} R_{0,j}^{-\alpha_d} |g_{0,j}|^2\Big)\bigg]\right]\nonumber \\
&\eqtop{a} \exp\left( -2 \pi \lambda_d \int_0^{\infty} \bigg(1-\mathbb{E}_{G}\bigg[\exp\Big(-\frac{\beta_d}{R_{0,0}^{-\alpha_d}} r^{-\alpha_d} G\Big)\bigg]\bigg)r\,\mathrm{d}r \right) \nonumber \\
&\eqtop{b}\exp\left(-2 \pi \lambda_d  \int_0^{\infty} \frac{r}{\frac{R_{0,0}^{\alpha_d}}{\beta_d} r^{\alpha_d}+1}\mathrm{d}r \right)\nonumber \\
&\eqtop{c} \exp\left(-\frac{\pi \lambda_d}{\mathrm{sinc}(\frac{2}{\alpha_d})}\Big(\frac{\beta_d}{R_{0,0}^{-\alpha_d}}\Big)^{2/\alpha_d}\right),
\label{eq:LT_I_d0}
\end{align}
where $(a)$ is based on the probability generating functional (PGFL) \cite{NET-032}, and $(b)$ follows from the fact that $G \sim \exp(1)$ and $\mathcal{L}\big[e^{-t}\big] = \frac{1}{s+1}$. Step $(c)$ follows by solving the integral in step $(b)$ and using $\mathrm{sinc}(x) = \frac{\sin(\pi x)}{\pi x}$.

Substituting \eqref{eq:LT_I_BS0} and \eqref{eq:LT_I_d0} in \eqref{eq:Pd2dProof}  concludes the proof of Proposition~\ref{proposition:P_succ_d2d}.\QEDA

\section{Proof of Proposition~\ref{proposition:P_succ_cue}}
\label{sec:proof_Pcov_cue}

Substituting $\mathrm{SINR}_c$ from \eqref{eq:SINR_CUE} into \eqref{eq:PsuccDef}, we get
\begin{align}
\mathrm{P}^c_{\mathrm{cov}}(\beta_c) &= \mathrm{Pr}\left\{ |\vect{h}_{0}^H \vect{v}_{0}|^2 \geq \frac{A_d}{\zeta}  D_{0,\textrm{BS}}^{\alpha_c}\Big(I_{d,c} + \frac{N_0}{A_d }\Big)\beta_c \right\} \nonumber\\
&\eqtop{a} \mathbb{E}_{D_{0,\textrm{BS}}, I_{d,c}}\left[e^{-\frac{A_d}{\zeta} D_{0,\textrm{BS}}^{\alpha_c}(I_{d,c} + \frac{N_0}{A_d})\beta_c }  \sum_{k=0}^{T_c - U_c} \frac{1}{k!}\bigg(\frac{A_d}{\zeta} D_{0,\textrm{BS}}^{\alpha_c}\Big(I_{d,c} + \frac{N_0}{A_d}\Big)\beta_c\bigg)^k\right] \nonumber \\
&\eqtop{b} \mathbb{E}_{D_{0,\textrm{BS}}, I_{d,c}}\left[e^{-\frac{N_0}{A_d}s} \sum_{k=0}^{T_c - U_c} \frac{s^k}{k!} \sum_{i=0}^{k} \binom{k}{i} \left(\frac{N_0}{A_d }\right)^{k-i} I_{d,c}^i e^{-s I_{d,c}} \right] \nonumber \\
&\eqtop{c} \mathbb{E}_{D_{0,\textrm{BS}}}\left[e^{-\frac{N_0}{A_d}s} \sum_{k=0}^{T_c - U_c} \frac{s^k}{k!} \sum_{i=0}^{k} \binom{k}{i} \left(\frac{N_0}{A_d }\right)^{k-i} \mathbb{E}_{I_{d,c}}\Big[I_{d,c}^i e^{-s I_{d,c}} \Big] \right]\nonumber\\
&\eqtop{d} \mathbb{E}_{D_{0,\textrm{BS}}}\Bigg[e^{-\frac{N_0}{A_d}s} \sum_{k=0}^{T_c - U_c} \frac{s^k}{k!} \sum_{i=0}^{k} \binom{k}{i} \left(\frac{N_0}{A_d }\right)^{k-i} (-1)^i\frac{\mathrm{d}^i}{{\mathrm{d}s}^i}\mathcal{L}_{I_{d,c}}(s)\Bigg], \label{eq:proof_Prob_cov_CUE}
\end{align}
where $(a)$ follows from the CCDF of $|\vect{h}_{0}^H \vect{v}_{0}|^2$ with $2|\vect{h}_{0}^H \vect{v}_{0}|^2 \sim \chi^2_{2(T_c-U_c+1)}$ given $D_{0,\textrm{BS}}$ and $I_{d,c}$. %
In $(b)$, we use Binomial expansion as
\begin{align}
\Big(I_{d,c} + \frac{N_0}{A_d}\Big)^k  &= \sum_{i=0}^k \binom{k}{i} \left(\frac{N_0}{A_d}\right)^{k-i} I_{d,c}^i,
\end{align}
and $(c)$ follows by taking the expectation with respect to the interference $I_{d,c}$. Step $(d)$ follows from
\begin{align}
\mathbb{E}_{I_{d,c}}\Big[I_{d,c}^i e^{-s I_{d,c}} \Big] = (-1)^i \frac{\mathrm{d}^i}{{\mathrm{d}s}^i} \mathcal{L}_{I_{d,c}}(s),\label{eq:E_Idc}
\end{align}
where $\mathcal{L}_{I_{d,c}}(s)$ is obtained using similar steps as in the derivation of $\mathcal{L}_{I_{d,0}}$ in \eqref{eq:LT_I_d0}:
\begin{align}
 \mathcal{L}_{I_{d,c}}(s) =
\exp\left(-\frac{\pi \lambda_d P_d^{2/\alpha_d}}{\mathrm{sinc}(\frac{2}{\alpha_d})} s^{2/\alpha_d}\right).\label{eq:L_dc}
\end{align}
Substituting \eqref{eq:L_dc} in \eqref{eq:proof_Prob_cov_CUE} and using the Fa\`{a} di Bruno's formula for the $i$-th derivative of a composite function $f(g(s))$ with $f(s) = e^s$ and $g(s) = - \frac{\pi \lambda_d P_d^{2/\alpha_d}}{\mathrm{sinc}(\frac{2}{\alpha_d})}  s^{2/\alpha_d}$, Proposition~\ref{proposition:P_succ_cue} follows.

~\QEDA

\bibliographystyle{IEEEtran}
\bibliography{IEEEabrv,refs,serveh}

\end{document}